\documentclass{emulateapj}
\usepackage{url}

\begin{document}

\shorttitle{Population II WD Masses}
\shortauthors{Kalirai et al.}

\title{The Masses of Population II White Dwarfs\altaffilmark{1,2,3}}

\author{
Jason S.\ Kalirai\altaffilmark{4}, D.\ Saul Davis\altaffilmark{5}, Harvey B.\ 
Richer\altaffilmark{5}\\ \vspace{0.2cm}
P.\ Bergeron\altaffilmark{6},
Marcio Catelan\altaffilmark{7,8},
Brad M.~S.\ Hansen\altaffilmark{9}, and
R. Michael Rich\altaffilmark{9},
}
\altaffiltext{1}{Data presented herein were obtained at the W.\ M.\ Keck
Observatory, which is operated as a scientific partnership among the
California Institute of Technology, the University of California, and the
National Aeronautics and Space Administration.  The Observatory was made
possible by the generous financial support of the W.\ M.\ Keck Foundation.}
\altaffiltext{2}{Based on observations obtained at the Gemini Observatory (Program 
IDs: GS-2005A-Q-5 and GS-2006B-Q-25), which is operated by the Association of Universities for Research in 
Astronomy, Inc., under a cooperative agreement with the NSF on behalf of the Gemini 
partnership: the National Science Foundation (United States), the Particle 
Physics and Astronomy Research Council (United Kingdom), the National Research 
Council (Canada), CONICYT (Chile), the Australian Research Council (Australia), 
CNPq (Brazil) and CONICET (Argentina).}
\altaffiltext{3}{Based on observations with the NASA/ESA {\it Hubble Space 
Telescope}, obtained at the Space Telescope Science Institute, which is operated 
by the Association of Universities for Research in Astronomy, Inc., under NASA 
contract NAS5-26555.  These observations are associated with proposal GO-8679.}
\altaffiltext{4}{Space Telescope Science Institute, 3700 San Martin Drive, Baltimore, 
MD, 21218; jkalirai@stsci.edu}
\altaffiltext{5}{Department of Physics and Astronomy, University of British Columbia, 
Vancouver, British Columbia, Canada, V6T~1Z1; sdavis/richer@astro.ubc.ca}
\altaffiltext{6}{D\'epartement de Physique, Universit\'e de Montr\'eal,
 C.P.~6128, Succ.~Centre-Ville, Montr\'eal, Qu\'ebec, Canada, H3C 3J7; bergeron@astro.umontreal.ca}
\altaffiltext{7}{Departamento de Astronom\'{i}a y Astrof\'{i}sica, Pontificia Universidad
Cat\'olica de Chile, Av. Vicu\~na Mackenna 4860, 782-0436 Macul, Santiago, Chile; mcatelan@astro.puc.cl}
\altaffiltext{8}{John Simon Guggenheim Memorial Foundation Fellow}
\altaffiltext{9}{Department of Physics and Astronomy, Box 951547, Knudsen Hall, 
University of California at Los Angeles, Los Angeles CA, 90095; hansen/rmr@astro.ucla.edu}


\begin{abstract}
Globular star clusters are among the first stellar populations 
to have formed in the Milky Way, and thus only a small sliver of their initial 
spectrum of stellar types are still burning hydrogen on the main-sequence today.  
Almost all of the stars born with more mass than 0.8~$M_\odot$ have evolved to 
form the white dwarf cooling sequence of these systems, and the distribution and 
properties of these remnants uniquely holds clues related to the nature of 
the now evolved progenitor stars.  With ultra-deep HST imaging observations, 
rich white dwarf populations of four nearby Milky Way globular clusters 
have recently been uncovered, and are found to extend an impressive 5 -- 8 
magnitudes in the faint-blue region of the H-R diagram.  In this paper, we 
characterize the properties of these population II remnants by presenting the 
first direct mass measurements of individual white dwarfs near the tip of the 
cooling sequence in the nearest of the Milky Way globulars, M4.  Based on 
Gemini/GMOS and Keck/LRIS multiobject spectroscopic observations, our results indicate that 
0.8~$M_\odot$ population II main-sequence stars evolving today form 0.53 $\pm$ 
0.01~$M_\odot$ white dwarfs.  We discuss the implications of this result as it 
relates to our understanding of stellar structure and evolution of population II 
stars and for the age of the Galactic halo, as measured with white dwarf 
cooling theory.
\end{abstract}

\keywords{globular clusters: individual (M4) -- stars: evolution -- stars: mass loss 
-- techniques: spectroscopic -- white dwarfs}


\section{Introduction} \label{introduction}

The end products of stellar evolution for 98\% of all stars will be white dwarfs.  
This final state results because most stars are not massive enough to ignite C 
and O, so that nuclear reactions cease with the formation of a degenerate core of 
He or a combination of He, C, and O.  With the cessation of nuclear reactions, 
white dwarfs emit light only via their cooling and become dimmer as time passes.  
These remnants have very thin surface layers of H and/or He, the bulk of the 
envelope having been blown away during relatively quiescent stages of mass loss 
in the progenitor star.  For the oldest stellar populations, 
$>$15\% of the entire present day mass of the system is tied up in white dwarfs, 
and therefore their distribution and detailed properties represent a unique link 
to explore the nature of the now-evolved population II progenitors.  Such studies 
are enhanced in star clusters where all of the remnants are co-spatial, have 
the same total age, and evolved from stars with the same metallicity, yet over 
a range in initial mass.


\begin{figure*}[ht]
\begin{center}
\leavevmode 
\includegraphics[width=13.0cm, angle=270]{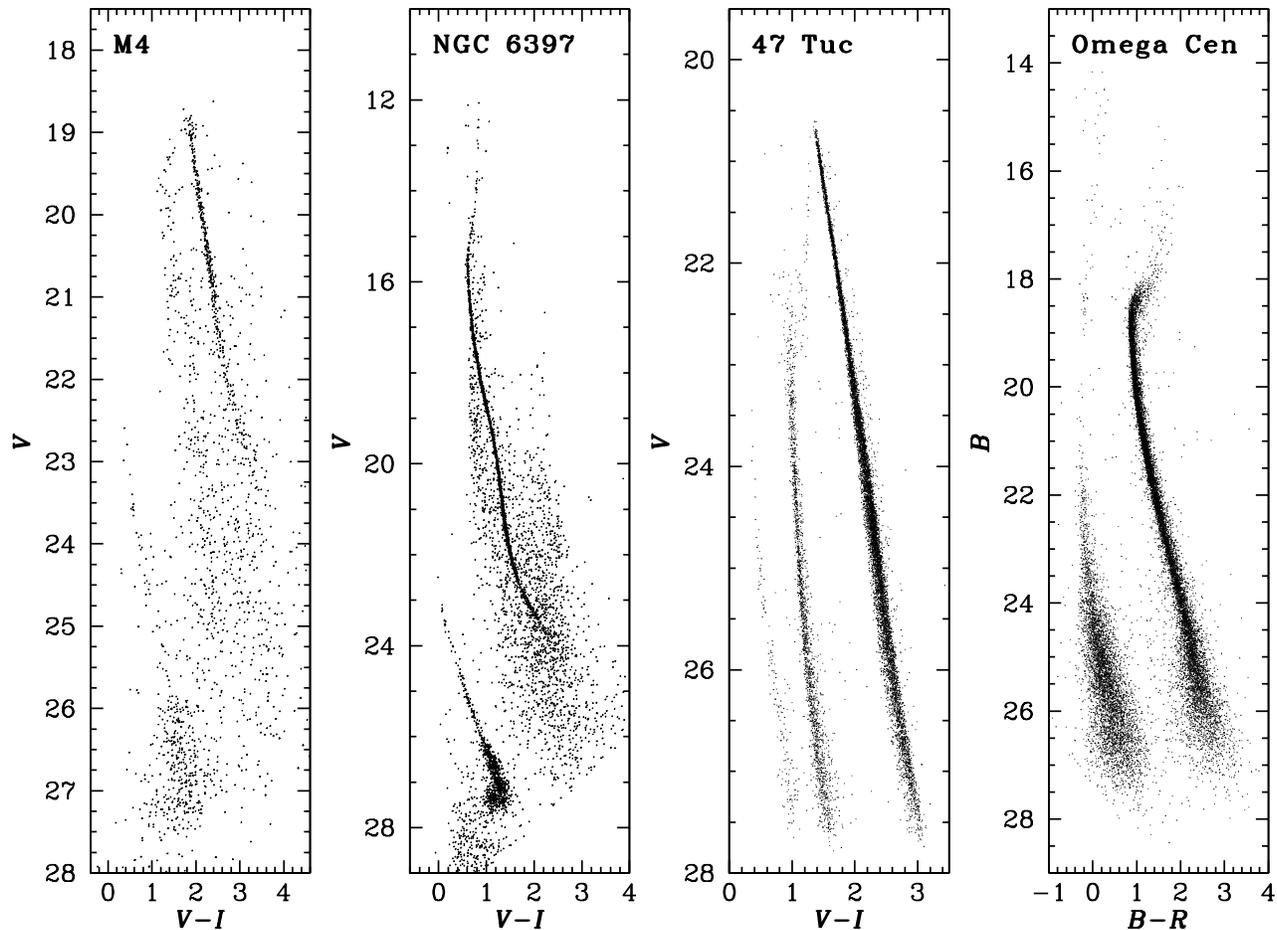}
\end{center}
\caption{HST CMDs of four nearby globular star clusters, extending 
from bright giant stars to the faintest white dwarfs (M4 -- Richer et~al.\ 2004, 
NGC~6397 -- Richer et~al.\ 2008, 47~Tuc -- J. Anderson, 2009, private 
communication, Omega~Cen -- Calamida et~al.\ 2008).  The faint-blue 
remnants in each cluster represent evolution of stars that were more 
massive than the present day turnoff, $M$ $\sim$ 0.8~$M_\odot$.  As 
discussed in \S\,\ref{introduction}, the distribution and properties of 
these white dwarfs hold important clues for our understanding of stellar 
evolution processes in population II systems. \label{fig:globsWDs}}
\end{figure*}


Milky Way globular star clusters have been dated to a formation time roughly 
12~Gyr ago (e.g., Krauss \& Chaboyer 2003), and therefore represent the first 
structures to form in the Galaxy.  
These systems are one of the most important tools that we have to probe the 
formation and evolution processes of galaxies \citep{brodie06}.  Given their 
ages, the present day main-sequence turnoff mass in globulars is about 
0.8~$M_\odot$, and therefore the bulk of their population has evolved through 
post main-sequence evolution, the majority having formed faint white dwarf stars.  
Through almost 400 orbits of Hubble Space Telescope imaging observations this 
decade, our team has uncovered the complete white dwarf cooling sequences of 
the two nearest globular clusters, M4 and NGC~6397 \citep{richer04,richer08}, 
and will target the third nearest cluster, 47~Tuc, in Cycle 17 (GO-11677).  
A shallower study of the massive globular Omega~Cen by \cite{monelli05} and 
\cite{calamida08} has successfully uncovered over 6500 remnants in 
this single cluster\footnote{Given the larger distance, the brightest faint-blue 
objects on the apparent Omega~Cen white dwarf cooling sequence may be subdwarf 
stars.}, roughly a quarter of the total known white dwarf population in the 
entire Milky Way from all other studies combined (e.g., Eisenstein et~al.\ 
2006, and subsequent Sloan Digital Sky Survey work).  The color-magnitude diagrams 
(CMDs) of these clusters, extending to impressively faint magnitudes, are presented 
in Figure~\ref{fig:globsWDs} (see also Bedin et~al.\ 2009 for a more recent ACS 
study of M4).

The white dwarf cooling sequences of nearby globular clusters have been 
analyzed to independently (e.g., of main-sequence stars and turnoff morphology) 
derive the distances and ages of the clusters \citep{hansen04,hansen07}, 
explore age variations between metal-rich and metal-poor systems using stars 
whose evolution is almost completely independent of metallicity \citep{hansen07}, 
possibly suggest large fractions of helium-core white dwarfs produced through binary 
interactions \citep{monelli05,calamida08}, and hint that mass loss stages leading to the 
formation of white dwarfs may involve small asymmetric kicks \citep{davis08}.  
For each of these studies, the interpretation of the photometry to yield the 
results depends on knowledge of the mass of the brightest white dwarfs that are 
forming today in these old systems.  Yet, at present, {\it no direct measurements of 
the masses of such population II white dwarfs exists}, although several indirect arguments 
suggest that the value should be $M_{\rm final}$ $\sim$ 0.53~$M_\odot$.  As discussed 
in \cite{renzini88} and \cite{renzini96}, stellar evolution theory and our understanding of the 
maximum luminosity of red giant (RG) and asymptotic giant branch (AGB) stars, 
as well as the luminosity of the horizontal branch (HB), predict that white 
dwarfs forming in globular clusters today should have 0.51 $ < M_{\rm final} <$ 
0.55~$M_\odot$.  The best observational verification of this theoretical 
prediction in the case of cluster white dwarfs was provided recently by \cite{moehler04}, 
who reported the masses of white dwarfs in NGC~6752 to be 0.53~$M_\odot$.  However, 
their spectroscopic measurements were only of sufficient quality to yield the temperatures 
of the white dwarfs, and therefore the photometry from the CMD was leveraged to derive 
masses.  Recently, a study of the masses of field white dwarfs was also presented 
from the SPY project (ESO SN Ia Progenitor surveY).  Pauli et~al.\ (2006) find several 
hundred thin disk and thick disk white dwarfs, and seven candidate halo objects based 
on kinematic analysis.  The masses of six of these white dwarfs range from 0.44 -- 0.51~$M_\odot$, 
with an additional lower mass star at 0.35~$M_\odot$.

In this paper we present the first results from a new study aimed at {\it directly} 
(i.e., purely from spectroscopic absorption line fitting) measuring the masses of known 
population~II white dwarfs in different environments.  Our goal is to target the brightest white 
dwarfs in the nearest globular star clusters (such as M4, NGC~6397, and 47~Tuc) and 
in the halo field, with ground-based spectroscopic instruments on 8--10~m telescopes.  
The clusters listed above span an appreciable range in metallicity (factor of 20) and 
are located at distances such that the brightest white dwarfs with $M_V$ = 10.5 have 
apparent magnitudes of $V$ = 22 -- 24.  The faintness of these stars, crowding 
issues in the cluster, and the need to obtain high signal-to-noise (S/N) spectra of the 
higher order Balmer lines at $\lambda <$ 4000~${\rm \AA}$ make this a difficult, but 
possible, project with current telescopes.   Below, we discuss the results from 
our first study, in which we target two dozen white dwarfs in M4 with the Gemini/GMOS 
and Keck~I/LRIS multiobject spectrographs over several observing runs.  The spectra 
for six of these stars, that are confirmed as cluster members and that have sufficient 
S/N to yield accurate mass measurements, are consistent with $M_{\rm final}$ = 0.53 $\pm$ 
0.01~$M_\odot$.  The impact of this measurement for bettering our knowledge of stellar 
structure and evolution, and for dating halo globular clusters through white dwarf 
cooling theory, are discussed.  An analysis of the remaining M4 spectra that did not yield 
masses, e.g., to define the spectral types of population II white dwarfs, is presented in 
the companion paper by \cite{davis09}.


\section{Imaging and Spectroscopic \\ Observations of M4} \label{observations}

The deep HST imaging data of M4 illustrated in Figure~\ref{fig:globsWDs} (left) were 
collected with the Wide Field Planetary Camera 2 (WFPC2) camera in 2001 as a part 
of GO-8679 \citep{richer04}.  The 
observations spanned a total of 123 orbits, in the broadband $F606W$ 
and $F814W$ filters, and were centered on a field located at ($\alpha$,$\delta$) 
= ($16^{\rm h}23^{\rm m}54.6^{\rm s}$, $-26^{\rm \circ}32{'}24.3{''}$), 
2.4~pc (4.8$'$) S-E of the center of the cluster.  Although the HST photometry 
is very accurate, and there is no ambiguity on the membership of the white 
dwarfs, the bulk of the stars are too faint to be useful for spectroscopic 
follow up with modern day facilities.  Additionally, the WFPC2 field of view is 5.3~square 
arcminutes, about six times smaller than our ground based instruments, and therefore 
the spectroscopic program would be highly inefficient with just this input list of 
targets.  Below we first discuss supplemental imaging of these data with the Gemini 
South telescope (\S\,\ref{geminiimaging}), followed by spectroscopic follow up 
(\S\,\ref{geminispect}).  In \S\S\,\ref{keckspect1} and \ref{keckspect}, we discuss 
additional spectroscopic observations of M4's white dwarf population with 
the Keck~I telescope.

\subsection{Gemini/GMOS Photometry and Astrometry} \label{geminiimaging}

\begin{figure}
\includegraphics[width=8.35cm]{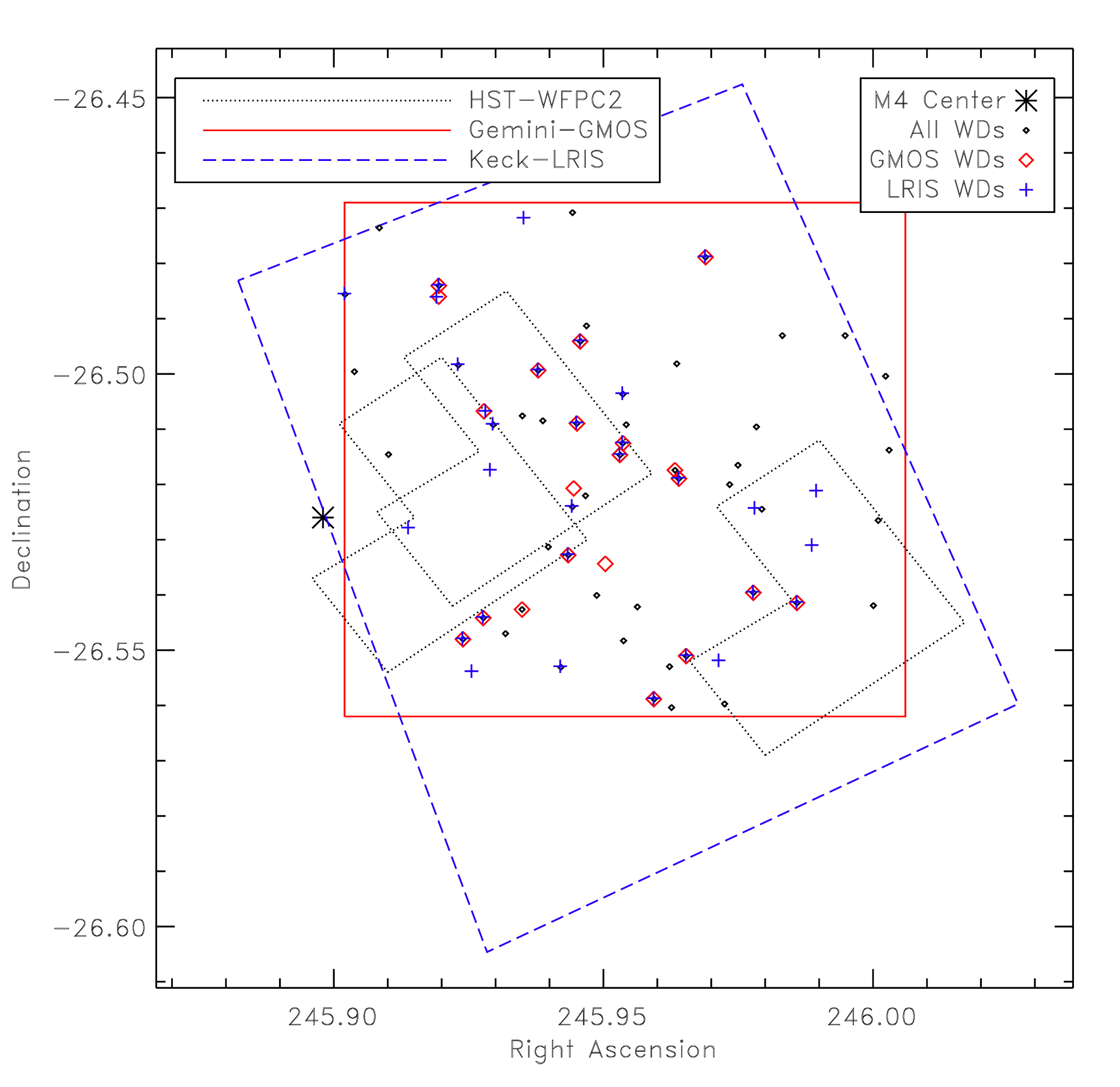}
\caption{The footprint of the HST/WFPC2 (dotted line), Gemini/GMOS (solid line), and 
Keck/LRIS (dashed) instruments.  The Cycle 9 deep HST observations shown in Figure~1 
comprise the bottom-right pointing, with the two other HST pointings being archive 
observations that we retrieved.  The white dwarf candidates selected from the GMOS 
photometry (i.e., those stars with magnitudes between $V$ = 20 and 24, and colors 
less than $V-I \sim$ 0.5) are shown as dots.  The objects targeted for GMOS 
spectroscopy are shown as diamonds and those for LRIS spectroscopy are represented 
with pluses.  The center of M4 is located at ($\alpha$, $\delta$) = 
(254.898$^\circ$,$-$26.526) and is indicated with an asterisk. \label{fig:foot}}
\end{figure}

We reobserved M4 with the Gemini South telescope and Gemini MultiObject Spectrograph 
(GMOS), anchoring the deep HST field near the south-eastern edge of the ground based 
image and placing shallower HST pointings near the western side.  The GMOS photometry, over a 5.5$' 
\times$ 5.5$'$ area, has the added advantage that it extends continuous imaging into the 
denser region towards the center of M4 and therefore provides a higher density of targets 
down to a given magnitude limit.  Of course, crowding in the ground-based data prevents 
the targeting of white dwarfs near bright stars, and this increases in the direction of 
the cluster center.  A footprint of the HST and GMOS images is provided in 
Figure~\ref{fig:foot}.


As this map illustrates, less than half of the 
Gemini/GMOS field of view overlaps the HST/WFPC2 observations.  In order 
to effectively select white dwarf candidates over the entire GMOS 
field, we obtained imaging in both the $g$$^\prime$ and $r$$^\prime$ filters, 
down to a depth of $\sim$25th mag.  The images were corrected for 
bias and flat fielding by the Gemini {\sc iraf} pipeline, version 1.4.  
The processed images were then reduced with the standard DAOPHOT/ALLSTAR 
photometry package to yield PSF photometry \citep{stetson94}.  Further 
details on the steps involved in our application are provided in 
\cite{kalirai08} for a similar study.

The matched catalog of WFPC2 and GMOS sources contains $10^3$ stars.  
We calibrate the $g$$^\prime$ and $r$$^\prime$ magnitudes to HST $F555W$ and $F814W$ 
(hereafter called $V$ and $I$) using derived transformations for the common stars in the 
space- and ground-based systems (see Davis et~al.\ 2009 for details on the 
transformations).  These offsets are then applied to the entire GMOS catalog.  
The positions of all stars on the GMOS image were carefully measured with respect 
to both HST positions and standard USNO guide stars to yield precise astrometry. 


\begin{figure}
\includegraphics[width=8.5cm]{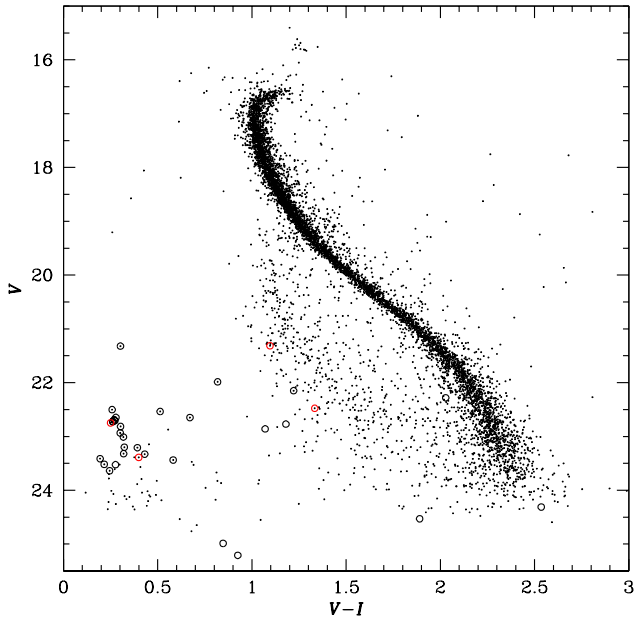}
\caption{The CMD constructed from the GMOS photometry (small dots).  There 
is a clearly identified white dwarf cooling sequence extending from $22 < V < 24.5$
with an approximate color of $V-I$ = 0.3 (one brighter object is also present).  
This CMD is contaminated by a more distant field population, located blueward of the main 
sequence.  There is also contamination in the white dwarf region of the 
CMD from field white dwarfs and possibly blue compact galaxies, which we address 
in \S\,\ref{membership}.  The open circles encircling the dots represent those 
objects which were targeted by Keck/LRIS for spectroscopy (see \S\,\ref{keckspect}).  
The red open circles are additional objects that were targeted by Gemini/GMOS, and 
that are not in the Keck sample due to slit conflicts (see \S\S\,\ref{geminispect} 
and \ref{keckspect}).  The open circles lacking small dots are HST detections for which we do 
not have ground based photometry (two of these stars are off the plot).  
These latter objects were selected for the sole purpose of filling the 
mask in relatively empty regions, and therefore no pre-selection was used.
\label{fig:cmd3}}
\end{figure}


The Gemini/GMOS CMD of M4 is presented in Figure~\ref{fig:cmd3}.  
Both the main-sequence and white dwarf cooling sequence are 
well measured, the latter containing about two dozen candidates 
brighter than $V$ = 24.  The location of these white dwarf candidates 
are illustrated on Figure~\ref{fig:foot} as smaller points.  The selection here 
includes stars with 20 $<$ $V$ $<$ 24.5 and colors $V-I$ $<$ 0.5.  Note 
the paucity of ground based candidates in the two inner WFPC2 fields (top-left).  
There are in fact many white dwarfs at these positions, however, the crowding 
limits their detection from the ground.

We further subdivide the sample of white dwarf candidates into two bins, with 
priority 1 stars being our strongest white dwarf candidates in the dominant 
part of the cooling sequence, and priority 2 objects being somewhat more uncertain 
objects that are located off the cooling sequence (e.g., redder or bluer) and/or 
near a bright star.  These latter objects are only included in the spectroscopic 
mask if they do not overlap a priority 1 star (hence, we get a spectrum of the 
potential white dwarf for free).  Any of the confirmed members of M4, based on 
HST proper motions in the region of overlap, were also targeted as priority 1 
objects.  Finally, we include several redder stars that are candidate CVs in 
the priority 2 sample; most of these are HST proper motion members as well. 
In total, 21 white dwarf candidates were targeted with Gemini/GMOS for follow 
up spectroscopy.  Note, a few of the very red or faint objects shown in the CMD 
were included without pre-selection for the purpose of filling the spectroscopic 
mask (these objects were not detected from the ground).

An example of the GMOS and WFPC2 photometry, astrometry, and a postage stamp 
image of the white dwarf candidate WD~09 is shown in Figure~\ref{fig:l09}.  
Similar figures for all other white dwarf candidates are provided in the 
Appendix.


\begin{figure}
\includegraphics[width=8cm]{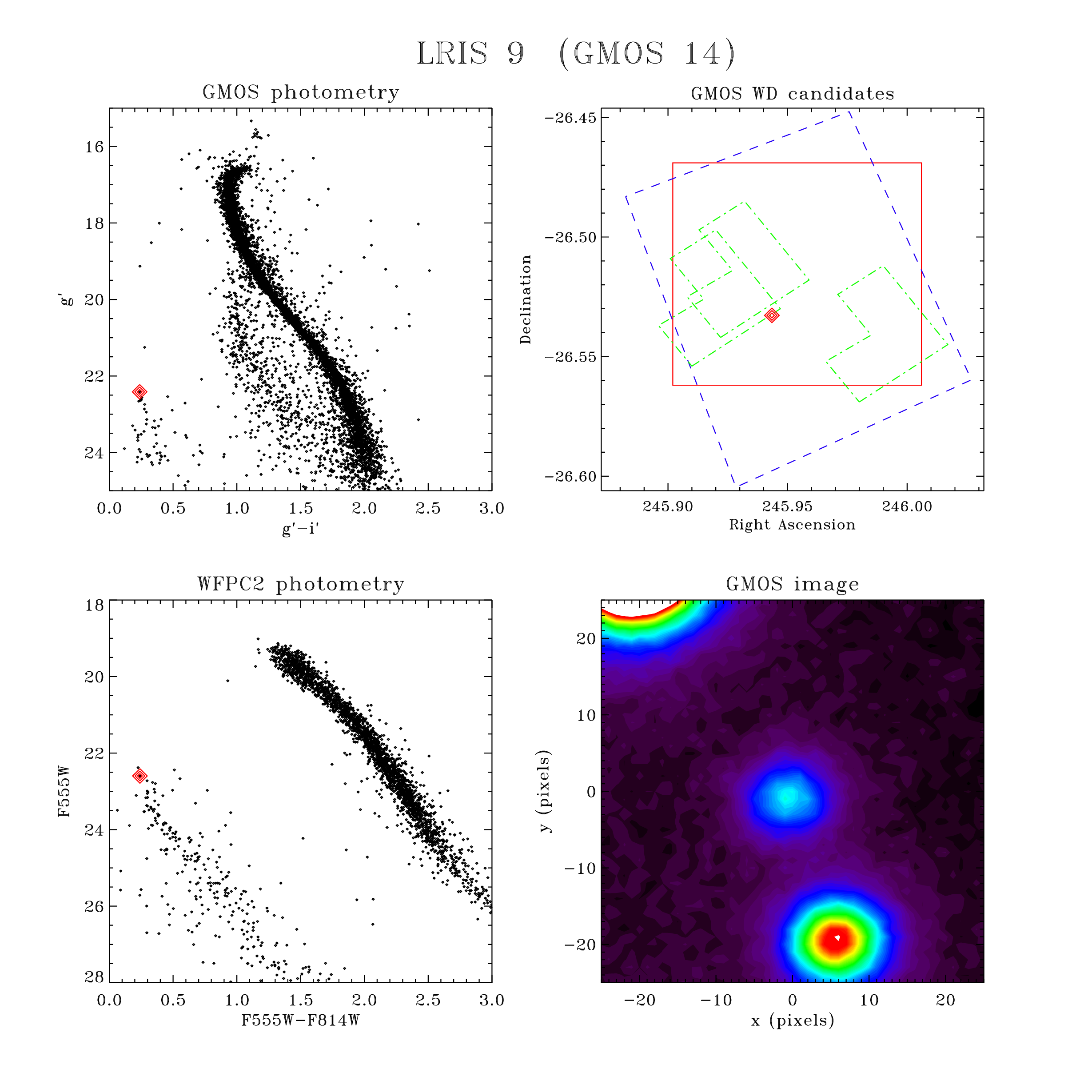}
\caption{The GMOS photometry (upper left), WFPC2 photometry (lower
left), astrometry showing position relative to the instrumental
footprints (upper right), and postage stamp section of the GMOS image 
centered on the target (lower right) for object WD~09.
\label{fig:l09}} 
\end{figure}


\subsection{Gemini/GMOS Spectroscopy} \label{geminispect}

In total, we obtained approximately 14.5 and 9 hours of queue scheduled 
spectroscopic science exposures with Gemini/GMOS in June/August 2005 and August/September 2006, 
respectively. The observations were taken with the B600 grating, covering a 
broad wavelength region of $\sim$2700~${\rm \AA}$ centered at 4680~${\rm \AA}$.  
The data were binned by a factor of two in the spectral direction to improve 
the S/N.  All science exposures were obtained in sub-arcsecond 
seeing conditions and photometric skies.

The raw data frames were downloaded from the Canadian Astronomy Data Center (CADC) 
in multi-extension FITS (MEF) format and reduced with the Gemini 
{\sc iraf} Package, version 1.4.  The reductions are described in detail in 
\cite{davis09}.  Of the 21 targeted white dwarfs, a spectrum was recovered for all but three.  
Remarkably, all 18 of the recovered spectra show pressure-broadened absorption 
lines indicative of white dwarfs.  The spectra for these 18 stars, as well as 
an analysis of their spectral types is also presented in Davis et~al.\ (2009).  
Unfortunately, in order to determine reliable spectroscopic masses, characterization 
of H$\epsilon$ is crucial, and even H$8$ (3889~${\rm \AA}$) is very useful.  The 
lack of flux at these wavelengths in our Gemini data limited the utility of the 
GMOS spectra for the central goals of this paper.  Although we could derive mass 
measurements from the spectra (see \S\,\ref{modelfit}), the uncertainties are several 
tenths of a solar mass and therefore we decided to supplement these data with 
higher S/N observations, especially at bluer wavelengths.


\subsection{Keck/LRIS Target Selection} \label{keckspect1}

With a sample of confirmed white dwarfs established from the GMOS observations, 
we applied for Keck~I telescope time to reobserve the targets using the 
LRIS multiobject spectrograph.  Not only does Keck provide a larger aperture 
than Gemini, but the spectrograph is very efficient in the blue and has an 
atmospheric dispersion corrector to minimize light loss at large airmasses (this is 
important for this study since Keck is in the north and M4 is in the south).  Given 
the similarities in the field of view (LRIS has a 5$' \times$ 7$'$ footprint), 
we simply used as many of the {\it confirmed} sample of white dwarfs from Gemini 
as our input, as well as several additional redder stars in the same brightness 
range.  The location of the 31 LRIS targets in the ground-based CMD are shown in 
Figure~\ref{fig:cmd3} as larger open circles.  The objects marked in red circles 
are those that were targeted with Gemini, but not Keck.  The LRIS footprint 
is also shown in Figure~\ref{fig:foot}.

\subsection{Keck/LRIS Spectroscopy} \label{keckspect}

Keck/LRIS spectroscopic observations (R.~M.\ Rich, PI) of the single M4 mask 
proved challenging.  Half nights, required due to the low declination of M4, 
were scheduled in 2005 June, 2007 April/July, and 2008 April, with the 2007 and 
2008 observations benefiting from the LRIS atmospheric dispersion correction 
system.  The multiple allocations were required due to poor weather that resulted 
in the loss of a significant number of scheduled hours. Altogether, we obtained 
10.6 hours of useful exposure time.  Even these data are of highly variable quality, 
with seeing measurements ranging from 0\farcs8 to over 2\farcs0.  In these crowded 
fields, our final spectra are dominated by a few of the contributions from the best 
nights.

LRIS \citep{oke95} has a dichroic that splits the spectrum into two channels at 
$\sim$5500~${\rm \AA}$. On the blue side, we use the 400/3400 grism 
(dispersion = 1.09 ${\rm \AA}$/pixel) which covers a wavelength region 
from the atmospheric cutoff to the dichroic.  For the red side, we use 
the 600/7500 grating (dispersion = 1.28 ${\rm \AA}$/pixel), 
centered at 6600~${\rm \AA}$, which covers a wavelength baseline of 
2620~${\rm \AA}$.  At our resolution and central wavelength, only 
H$\alpha$, which is a rather poor mass indicator, landed on the red 
side.  Hence, we chose not to reduce the red-side data for this paper.

The blue-side LRIS spectra were reduced using standard {\sc iraf} tasks.  
The trace, sky subtraction, and extraction were all performed with the
{\sc apall} task.  As expected given the better blue sensitivity of LRIS 
compared to GMOS, the trace at blue wavelengths was far more certain. 
Given the high levels of crowding and background light from bright M4 
stars in the field, our ability to obtain a reliable trace, and perform 
accurate sky subtraction is dependent on the particulars of each individual 
slitlet (see \S\,\ref{culling}).

The wavelength calibration was calculated in {\sc iraf} from spectra of three 
lamps containing Hg, Zn, and Cd.  The typical final dispersion of the wavelength 
solution was well constrained, with uncertainties of $\sim$0.1~${\rm \AA}$.  The 
reduced, wavelength calibrated LRIS spectra were 
flux calibrated using the spectrophotometric standard HZ44.  Finally, the 
individual spectra were combined using {\sc scombine}, with weights assigned 
according to the individual S/N ratios.

We summarize the identification, positions, and brightnesses of all white 
dwarf candidates in Table~\ref{targtab.tab}.  The four stars that have 
slit conflicts, and therefore were not observed with Keck, are listed 
at the bottom of the table.  Altogether, 31 objects were targeted with 
Keck/LRIS.  Note, the identification of these stars as ``WD'' at this stage 
does not confirm their nature, although we will find later that most are in 
fact white dwarfs.  We use this naming scheme for convenience as 
we will refer back to this Table in future sections.


\begin{deluxetable}{lcccc}
\small
\tablecaption{\label{targtab.tab}Properties of the M4 White Dwarf Targets}
\tablecolumns{5}
\tablehead{\colhead{Identification \#} &
           \colhead{RA (J2000)} &
           \colhead{DEC (J2000)} &
           \colhead{$V-I$} &
           \colhead{$V$}}
\startdata
WD~00 & 245.9579 & $-$26.5589 & 0.32 & 23.32     \\ %
WD~01 & 245.9256 & $-$26.5540 & 1.18 & 22.77     \\ %
WD~02 & 245.9406 & $-$26.5531 & 0.30 & 21.32 \\
WD~03$^{\rm *}$ & 245.9710 & $-$26.5511 & 0.93 & 25.21 \\ %
WD~04 & 245.9638 & $-$26.5511 & 0.27 & 22.69   \\ %
WD~05 & 245.9224 & $-$26.5480 & 0.27 & 22.71    \\ %
WD~06 & 245.9262 & $-$26.5442 & 0.28 & 22.65   \\ %
WD~07 & 245.9844 & $-$26.5414 & 0.82 & 21.99   \\ %
WD~08 & 245.9763 & $-$26.5396 & 0.67 & 22.65  \\ %
WD~09 & 245.9419 & $-$26.5328 & 0.26 & 22.50  \\ %
WD~10$^{\rm *}$ & 245.9120 & $-$26.5281 & 2.03 & 22.28 \\ %
WD~11$^{\rm *}$ & 245.9867 & $-$26.5300 & 1.31 & 26.96 \\ %
WD~12 & 245.9427 & $-$26.5241 & 0.22 & 23.52  \\ %
WD~13$^{\rm *}$ & 245.9764 & $-$26.5240 & 2.54 & 24.31 \\ %
WD~14$^{\rm *}$ & 245.9890 & $-$26.5214 & 0.85 & 24.99 \\ %
WD~15 & 245.9625 & $-$26.5190 & 0.26 & 22.73 \\ %
WD~16$^{\rm *}$ & 245.9276 & $-$26.5175 & 0.28 & 23.53 \\ %
WD~17 & 245.9515 & $-$26.5147 & 0.32 & 23.20 \\ %
WD~18 & 245.9521 & $-$26.5125 & 0.25 & 23.64 \\ %
WD~19 & 245.9281 & $-$26.5092 & 0.43 & 23.33 \\ %
WD~20 & 245.9436 & $-$26.5090 & 0.32 & 23.01  \\ %
WD~21 & 245.9263 & $-$26.5067 & 1.22 & 22.15  \\ %
WD~22 & 245.9521 & $-$26.5037 & 0.19 & 23.41  \\ %
WD~23 & 245.9364 & $-$26.4994 & 0.51 & 22.54  \\ %
WD~24 & 245.9216 & $-$26.4984 & 0.26 & 22.72  \\ %
WD~25 & 245.9442 & $-$26.4941 & 0.39 & 23.21  \\ %
WD~26 & 245.9168 & $-$26.4860 & 1.89 & 24.53  \\ %
WD~27 & 245.9006 & $-$26.4856 & 0.58 & 23.44  \\ %
WD~28 & 245.9179 & $-$26.4840 & 0.30 & 22.81  \\ %
WD~29 & 245.9674 & $-$26.4789 & 0.30 & 22.94  \\ %
WD~30 & 245.9353 & $-$26.4719 & 1.07 & 22.86  \\ %
GemWD~05 & 245.9195 & $-$26.4860 & 1.33 & 22.48 \\ %
GemWD~11 & 245.9633 & $-$26.5174 & 0.40 & 23.39 \\ %
GemWD~15 & 245.9349 & $-$26.5426 & 0.25 & 22.75 \\ %
GemWD~17 & 245.9504 & $-$26.5344 & 1.10 & 21.31 \\  %

\enddata
\tablenotetext{*}{These stars were not detected from the ground; magnitudes are from the HST photometry.}
\normalsize
\end{deluxetable}



\begin{figure*}
\begin{center}
\leavevmode 
\includegraphics[width=13.0cm,angle=270]{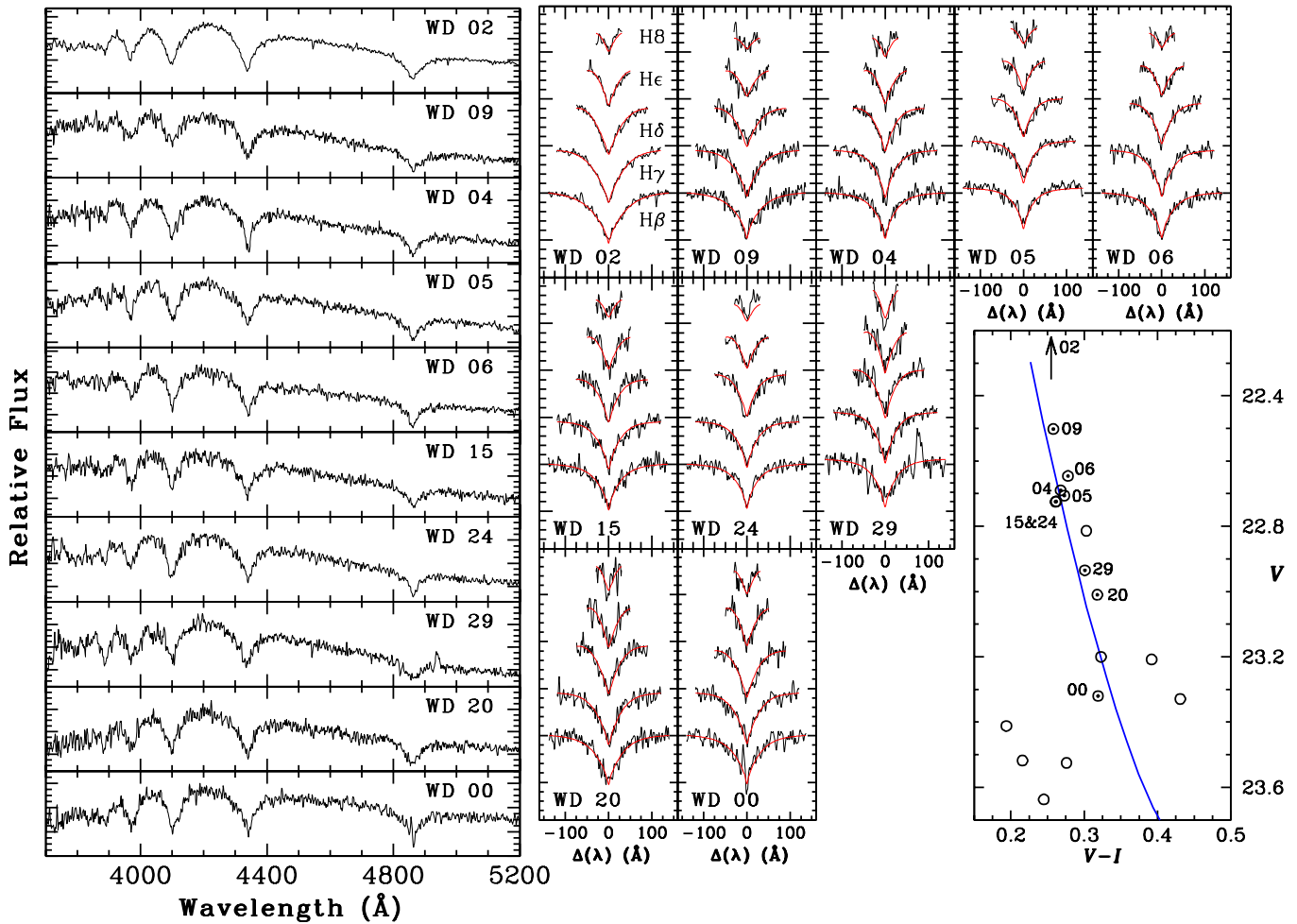}
\end{center}
\caption{Keck/LRIS spectra for ten white dwarfs along the sightline 
to M4 (left).  The S/N per resolution element at a wavelength 
of $\sim$4030~${\rm \AA}$ in these data is 30 -- 50; the higher order hydrogen 
Balmer lines are well characterized.  The spectra for each of these 
stars is fit to atmosphere models to yield $T_{\rm eff}$, log~$g$, 
and $M_{\rm final}$ for the white dwarfs as described in \S\,\ref{modelfit} 
(top-right).  In these smaller panels, the Balmer lines for a given white 
dwarf are displayed with H$\beta$ at the bottom, and successively higher 
order transitions on top.  The simultaneous fit to all Balmer lines in the 
best fit model is shown as a smooth red curve.  A closer view of the white 
dwarf cooling sequence of M4 is shown in the CMD in the bottom-right panel, 
with a 0.5 $M_\odot$ theoretical cooling model \citep{wood95}.  Note, all spectra in 
this figure have been smoothed by a three pixel boxcar function for clarity.
\label{fig:wdspectraM4}}
\end{figure*}



\section{Eliminating Non-White Dwarfs and Stars With Poor Spectra} \label{culling}

The ground-based CMD of our targets in Figure~\ref{fig:cmd3} shows that 
the dominant, blue white dwarf cooling sequence of M4 extends from 
$V$ = 22.5 to 23.5, and contains eleven stars that were spectroscopically observed 
with LRIS (note, the clump at $V \sim$  22.7, $V-I$ $\sim$ 0.3 contains five 
LRIS targets).  An additional target, WD~02, is almost one magnitude brighter 
than the family of fainter points and has $V$ = 21.3, another four stars 
are found fainter and bluer than the cooling sequence at $V$ $\sim$ 23.5, 
$V-I$ $\sim$ 0.25 (including the one object without ground based photometry), 
and two stars are at the faint end and redder than the cooling sequence at
$V$ = 23.2 and $V-I$ = 0.4.  Most of the stars that are redder than this 
sample of 18 objects (i.e., $V-I$ $>$ 0.5) yielded spectra that are not 
interesting for the goals of this paper and will be commented upon further 
in \cite{davis09}.  In Table~1, these objects are WD~01, 03, 07, 08, 10, 11, 
13, 14, 21, 23, 26, 27, and 30.  The one exception to this is possibly the 
relatively bright star just beyond the $V-I$ = 0.5 color cut at $V$ = 22.54, $V-I$ = 0.51 
(WD~23).  This star is a proper motion member of M4 but, as shown in the Appendix in 
Figure~\ref{fig:wd18_26}, is located very close to a bright neighbor.  Likely, the 
color (and spectrum) of this star is contaminated from the neighboring star and 
we therefore eliminate it from our analysis.


\begin{table*}
\small
\begin{center}
\caption{}
\begin{tabular}{lcccccccc}
\hline
\hline
\multicolumn{1}{c}{ID} & \multicolumn{1}{c}{$V$} & \multicolumn{1}{c}{$T_{\rm eff}$ (K)$^{a}$} & 
\multicolumn{1}{c}{log~$g$$^{a}$} & \multicolumn{1}{c}{$M_{\rm final}$ ($M_\odot$)} & 
\multicolumn{1}{c}{$V_{\rm theory}$$^{b}$} & \multicolumn{1}{c}{$t_{\rm cool}$} & 
\multicolumn{1}{c}{S/N$^{c}$} & \multicolumn{1}{c}{Member$^{d}$} \\
\hline
WD~00       & 23.32 & 20,600 $\pm$ 600 & 7.75 $\pm$ 0.09 & 0.52 $\pm$ 0.04 & 22.82 $\pm$ 0.26 & 40  $\pm$  6 & 39 & YES (PM)  \\
WD~04       & 22.69 & 24,300 $\pm$ 500 & 7.68 $\pm$ 0.08 & 0.50 $\pm$ 0.03 & 22.40 $\pm$ 0.23 & 19  $\pm$  1 & 47 & YES (Lum) \\
WD~06       & 22.65 & 25,600 $\pm$ 500 & 7.87 $\pm$ 0.08 & 0.59 $\pm$ 0.04 & 22.59 $\pm$ 0.23 & 18  $\pm$  2 & 51 & YES (Lum) \\ 
WD~15       & 22.73 & 24,300 $\pm$ 600 & 7.79 $\pm$ 0.09 & 0.55 $\pm$ 0.04 & 22.57 $\pm$ 0.27 & 21  $\pm$  2 & 29 & YES (Lum) \\
WD~20       & 23.01 & 19,700 $\pm$ 600 & 7.73 $\pm$ 0.10 & 0.51 $\pm$ 0.05 & 22.89 $\pm$ 0.29 & 49  $\pm$  8 & 32 & YES (PM)  \\ 
WD~24       & 22.72 & 25,700 $\pm$ 500 & 7.70 $\pm$ 0.07 & 0.51 $\pm$ 0.03 & 22.33 $\pm$ 0.21 & 16  $\pm$ 1  & 54 & YES (PM)  \\  
WD~02       & 21.32 & 18,800 $\pm$ 200 & 8.16 $\pm$ 0.04 & 0.75 $\pm$ 0.03 & 23.56 $\pm$ 0.14 & 128 $\pm$ 11 & 61 & NO        \\
WD~05       & 22.71 & 28,200 $\pm$ 400 & 7.59 $\pm$ 0.08 & 0.48 $\pm$ 0.03 & 21.93 $\pm$ 0.22 & 12  $\pm$ 1  & 49 & NO        \\
WD~09       & 22.50 & 25,100 $\pm$ 600 & 8.19 $\pm$ 0.09 & 0.79 $\pm$ 0.05 & 23.10 $\pm$ 0.25 & 44  $\pm$ 12 & 32 & NO or DD  \\ 
WD~29       & 22.94 & 20,900 $\pm$ 700 & 7.44 $\pm$ 0.11 & 0.41 $\pm$ 0.04 & 22.32 $\pm$ 0.31 & 33  $\pm$  4 & 27 & NO        \\
\hline
\end{tabular}
\tablenotetext{$^{a}$}{The new Stark broadening calculations in \cite{tremblay09} have not been used to 
recalculate the $T_{\rm eff}$ and log~$g$ values (which would require approximate shifts of $+$1,500~K and 
0.1~dex), although the masses have been corrected as discussed in \S\,\ref{modelfit}.}
\tablenotetext{$^b$}{Theoretical luminosity from spectral fits (see \S\,\ref{membership}).  The error bar includes the 
1$\sigma$ error in the distance modulus (0.09 mags).}
\tablenotetext{$^c$}{The spectral S/N per resolution element, calculated between H$\delta$ and H$\epsilon$.} 
\tablenotetext{$^d$}{Membership evaluation (see \S\,\ref{membership}): PM = Proper Motion Member, Lum = Luminosity 
Member, DD = Double Degenerate Candidate.}
\label{tab:WDresults}
\end{center}
\end{table*}


The family of 18 white dwarf candidates in M4's 'pseudo' cooling sequence 
are all confirmed to be white dwarfs (see Davis et~al.\ 2009).  For our 
primary goal of measuring accurate masses, we want to first restrict this 
sample to only include well measured, isolated white dwarfs with robust 
S/N (see \S\,\ref{modelfit}).  The stars along the cooling 
sequence in our sample vary in brightness by about a factor of three, and 
therefore the fainter stars will intrinsically have much less signal than 
their brighter counterparts.  Added to this, we note that the final quality 
of the spectrum for any star will depend on several factors and therefore not 
always correlate with the brightness.  For example, the spatial length of the 
given slit effects the quality of the sky subtraction, the position of the star 
relative to M4's center effects the sky background level, and any astrometric 
mis-centering of the star within the slit causes light loss.  We first 
eliminate three (fainter) stars that are located within $\lesssim$30 pixels 
($\sim$2$''$) of a bright neighbor.  These stars, WD~16, 19, and 22, all 
exhibit a noisy spectrum that is likely contaminated with light from the 
nearby, brighter star (see Figures~\ref{fig:wd09_17} \& \ref{fig:wd18_26} in 
the Appendix).  We also eliminate five 
white dwarfs, four of which are isolated, for which our extractions resulted 
in S/N $<$ 10 spectra, WD~12, 17, 18, 25, and 28.  The extracted spectra for these 
white dwarfs suffer from unusable line profiles blueward of H$\delta$, but with improved 
S/N, these particular sources could also be analyzed as described below.


Our final sample has ten white dwarfs, 
all of which are close to the M4 cooling sequence and have $V <$ 23.4 (seven of 
which have $V <$ 23).  As we demonstrate in \S\,\ref{membership}, 
four of these ten white dwarfs, WD~00, 09, 20, and 24, are HST proper motion 
members of M4.  We present the spectra for these ten 
white dwarfs in Figure~\ref{fig:wdspectraM4} (left).   All of the spectra 
clearly show pressure-broadened Balmer lines extending from H$\beta$ to H8.  
These are therefore all DA (hydrogen) spectral type white dwarfs.  We report the 
spectral types of all of our white dwarfs, including stars that are too faint for 
mass constraints, in \cite{davis09}.  The S/N 
per resolution element in our ten stars is generally 40 -- 50, and at best 
60 for one star (WD~02).\footnote{The spectral S/N is calculated at 
$\sim$4030~${\rm \AA}$ as discussed in \S\,\ref{syntheticspectra}.}  

We also present a detailed view of the white dwarf cooling sequence in the CMD, 
and indicate the positions of these ten white dwarfs in the bottom-right 
panel of Figure~\ref{fig:wdspectraM4} (dotted circles).  The remaining open 
circles (without dots) in this diagram are those stars that were spectroscopically 
targeted in this narrow region of CMD space.  We note that the magnitudes and colors 
of WD~15 and 24 are almost identical and so the two dots can not be discerned.  
We also illustrate a theoretical white dwarf cooling model for $M$ = 0.5~$M_\odot$ 
\citep{wood95}, which cuts through our sample nicely.  From this plot, it is clear that WD~02 is 
well removed from the dominant sequence of points and the cooling model.  Although 
it is possible that the photometry for this star is in gross error, it is more 
likely that the star is a field white dwarf along M4's sightline (see 
\S\,\ref{membership} for more information).

\section{Mass Measurements of White Dwarfs in M4}

The technique of measuring masses of white dwarfs from Balmer line 
fitting has been widely employed for field stars, such as 
in the Palomar-Green (PG) Sample \citep{liebert05} and Sloan Digital 
Sky Survey (SDSS) \citep{eisenstein06,kepler07}.  The technique has also been 
cross-checked against independent measures of the mass of a white 
dwarf such as astrometry for binary systems (e.g., Sirius~B, Barstow 
et~al.\ 2005), fits to the mass-radius relation for stars with trigonometric 
parallaxes \citep{holberg08}, the gravitational redshift method 
\citep{reid96}, and pulsation mode analysis \citep{kawaler91}.  For 
temperatures hotter than $\sim$12,000~K, these methods are in excellent 
agreement with the spectroscopic line fitting technique \citep{bergeron95}.  

In this section, we first briefly describe our method to calculate masses 
for the ten white dwarfs discussed above.  We then verify the uncertainties in the masses 
using synthetic spectra of the same quality as the observations, and fit using 
the same methods described below.  Finally, we remove field white dwarfs from 
the sample and analyze the mass of the confirmed M4 members.

\subsection{Balmer Line Fits}\label{modelfit}

Fitting the Balmer lines of a white dwarf to model atmospheres involves 
reproducing multiple line profiles from H$\beta$ to higher order 
transitions \citep{bergeron92}.  Typically, four or five well measured lines are 
enough to constrain the temperature ($T_{\rm eff}$) and surface gravity 
(log~$g$) of the star \citep{bergeron92}.  The fifth Balmer line in this 
sequence, H$8$, is the weakest and bluest (at $\lambda$ = 3889~${\rm \AA}$), 
and therefore a spectrograph with high throughput in the blue is needed 
to characterize the feature accurately in faint stars.  General details 
of the fitting technique used to derive $T_{\rm eff}$ and 
log~$g$ are provided in \cite{bergeron92}, and recent refinements to 
this method are given in \cite{liebert05}.  Additional information on the 
fitting of white dwarf spectra similar to these observations, and taken with 
our instrumental setup, are presented in \S\,6 of \cite{kalirai08}.
Specifically, the improved method to normalize the line profiles described in 
\cite{liebert05} is used in our work.

Simultaneous fits of model atmospheres to all of the Balmer lines, in each 
of the ten white dwarfs discussed above, are presented in 
Figure~\ref{fig:wdspectraM4} (top-right).  Within each panel, the absorption 
lines of a given white dwarf are arranged with H$\beta$ at the bottom, and 
successive higher order transitions on top.  The model fits are shown as smooth 
red curves and are excellent in all cases.  We also confirmed that the two 
anomalous features, small spikes caused by bad sky subtraction in the H$\beta$ line 
of WD~00 and 29, do not affect the results of the fit.  This line is least sensitive 
to changes in the gravity of the star.  The output from the spectroscopic fits are 
the $T_{\rm eff}$ and log~$g$ for each star, which we summarize in Table~2.  

The masses of these ten stars ($M_{\rm final}$) are simply calculated by combining the 
measured surface gravity of the stars with the well established mass-radius relation for 
white dwarfs.  In general, most of the white dwarf masses are $\sim$0.5~$M_\odot$ with 
an uncertainty of $\lesssim$10\%.  We also estimate both the cooling ages, $t_{\rm cool}$, 
and the theoretical luminosity of each star by interpolating the $T_{\rm eff}$ and log~$g$ 
within evolutionary models similar to \cite{fontaine01}, but assuming a 50/50 
carbon-oxygen core mix.\footnote{The models are available at 
\url{http://www.astro.umontreal.ca/~bergeron/CoolingModels/.}}  Our default 
calculations assume thick hydrogen layers with $q(\rm H)$ = $M_{\rm H}/M$ = 10$^{-4}$ 
and helium layers of $q(\rm He)$ = 10$^{-2}$, and are summarized in Table~2 and 
discussed further below.  We also tested the sensitivity of our results to models 
(e.g., Wood 1995) with pure carbon cores and to models with very thin hydrogen 
layers (e.g., $q(\rm H)$ = 10$^{-10}$).  The results from these comparisons, for 
both the final remnant masses and the theoretical luminosities, are identical to 
our default values within the measured uncertainties in these properties (which 
are small, see Table~2).  We note, however, that the cooling ages of these 
bright, young white dwarfs are systematically larger in the pure C core models 
by 20 -- 30\%.

At the time of writing this paper, Tremblay \& Bergeron (2009) have just presented 
the first results from an improved calculation of the Stark 
broadening of hydrogen lines in dense plasmas that are representative of white dwarf 
atmospheres.  Their new results, including non-ideal effects, suggest that the masses 
measured through the spectroscopic technique require a slight upward revision of 
0.034~$M_\odot$.  As the temperatures of the stars using the new calculations are 
also slightly hotter (e.g., 1,000 -- 2,000~K), the luminosities are unaffected.  The 
full implications of this study have not yet been investigated and the dependency of 
this offset on mass is not well constrained. In the results that follow, we have 
adjusted our default masses by $+$0.034~$M_\odot$ to be consistent with these new 
calculations (including the reported masses in Table~2).  We have not 
corrected the individual $T_{\rm eff}$ and log~$g$ values by a common offset (which 
would be approximately 1,500~K and 0.1~dex, respectively).

\subsection{Testing the Uncertainties in $M_{\rm final}$: Synthetic Spectral Fits} 
\label{syntheticspectra}

The S/N of the ten white dwarfs in our sample is listed in column 8 
of Table~2.  This is calculated from the pseudo-continuum between the H$\delta$ and 
H$\epsilon$ Balmer lines, at $\sim$4030~${\rm \AA}$.  The pixel-to-pixel RMS 
scatter is converted to a S/N per resolution element by multiplying by the 
square root of the number of pixels per FWHM (six, in our case).  The bulk 
of the stars have S/N = 40 -- 50, although a few are as low as $\sim$30 and 
one object is above S/N = 60.  Note, these S/N values should only be used as 
a relative ranking of the quality of each spectra, and are not indicative of 
the global S/N.  We can see that, due to the throughput, the S/N 
decreases blueward of the wavelength at which we made this measurement.

\begin{figure}
\epsscale{1.1} \plotone{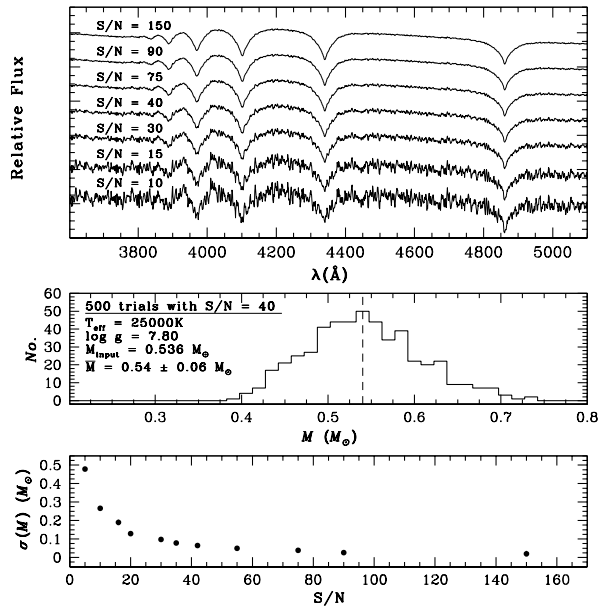} 
\figcaption{Synthetic white dwarf spectra are created over a range of S/N from 
10 to 150, as measured at $\sim$4030~${\rm \AA}$ between the H$\delta$ and 
H$\epsilon$ Balmer lines (top panel).  The typical white dwarf in our observations 
shown in Figure~\ref{fig:wdspectraM4} is similar in spectral quality to the S/N = 40 
case in this plot.  In the middle panel, we illustrate the results from re-fitting 
500 simulated spectra with this S/N.  The input spectra in this Monte Carlo trial 
all had $T_{\rm eff}$ = 25,000~K, log~$g$ = 7.80, and $M$ = 0.536~$M_\odot$.  Our 
results, using the same methods of fitting five Balmer lines as outlined 
in \S\,\ref{modelfit}, indicates that the mean recovered mass is $\langle$$M$$\rangle$ 
= 0.54~$M_\odot$ with a 1-$\sigma$ error in the mean of 0.06~$M_\odot$.  In the 
bottom panel, we illustrate the error in the recovered mass from similar trials of 
500 synthetic spectra for a series of S/N ratios.  For values comparable to our data, 
we should be able to recover the masses of these white dwarfs to $\sim$10\%.
\label{fig:synthetic}}
\end{figure}

In Figure~\ref{fig:synthetic} (top), we produce simulated white dwarf spectra over 
a range of S/N.  These spectra have been convolved with a Gaussian profile 
with 6~${\rm \AA}$ FWHM.  Directly comparing the quality of 
our data in Figure~\ref{fig:wdspectraM4} to these simulated spectra, we 
can conclude that most of our objects are better than the S/N = 30 case, 
and similar, if not better, than the S/N = 40 case.  For the latter, 
we generate a Monte-Carlo with 500 trials and fit each of these spectra in 
our fitting code.  The prescription used in the fitting is 
identical to that for our actual white dwarfs, namely, the same models 
are used in the fit with five Balmer lines.  The results from this 
test are shown in the middle panel of Figure~\ref{fig:synthetic}.  The input 
stellar parameters for the synthetic white dwarf 
are $T_{\rm eff}$ = 25,000~K, log~$g$ = 7.80, and $M$ = 0.536~$M_\odot$.  
From the distribution of recovered masses, we find $\langle$$M$$\rangle$ 
= 0.54~$M_\odot$ with a dispersion of 0.06 $M_\odot$.  Therefore, we 
expect to measure the masses of our white dwarfs to roughly 10\% 
precision based on this test, which agrees nicely with our reported 
uncertainties in mass (see Table~2, column 5).  
In the bottom panel of Figure~\ref{fig:synthetic}, we extend the test 
above for other S/N values and plot the associated error from this 
fitting.


\subsection{Establishing Cluster Membership} \label{membership}

Thus far, we have established masses for ten white dwarfs in the M4 
direction.  As we have reported in previous spectroscopic studies of white 
dwarfs in star clusters, some fraction of the stars along these lines of 
sight are actually members of the field population and therefore need to 
be removed before discussing the mass 
distribution of bonafide members.  The best method to establish such 
membership is through proper motion analysis, and we have such data 
from the HST overlap of a portion of our LRIS field (see Figure~\ref{fig:foot}).
Out of these ten white dwarfs, we can confirm that WD~00, 09, 20, and 24 
are proper motion members (as are WD~16 and 23; these are rejected from 
our sample based on the poor quality of the recovered spectra 
and/or contamination from a nearby neighbor -- see \S\,\ref{culling}).  This is 
not a surprising result since the proper motions were folded into the selection 
process to target white dwarfs in the region of overlap with our HST field.

In addition to these five confirmed M4 white dwarfs, we note that the 
remaining objects in our sample closely follow the 0.5~$M_\odot$ cooling 
sequence, and are consistent in position in the CMD with the confirmed 
members.  Although we do not have proper motions for these stars, some 
of them are likely members of M4.  To distinguish 
members from non-members, we can use the theoretical luminosity of each star 
to calculate its distance modulus.  Any stars that have significantly discrepant 
values from the known distance of M4, ($m-M$)$_V$ = 12.51 $\pm$ 0.09 
\citep{richer04}, should be eliminated from our sample as they likely 
represent field white dwarfs along the line of sight, binary white dwarfs, 
or possibly objects for which the spectroscopic fit is incorrect due to 
some unknown artifact in the spectrum.  


\begin{figure}
\epsscale{1.1} \plotone{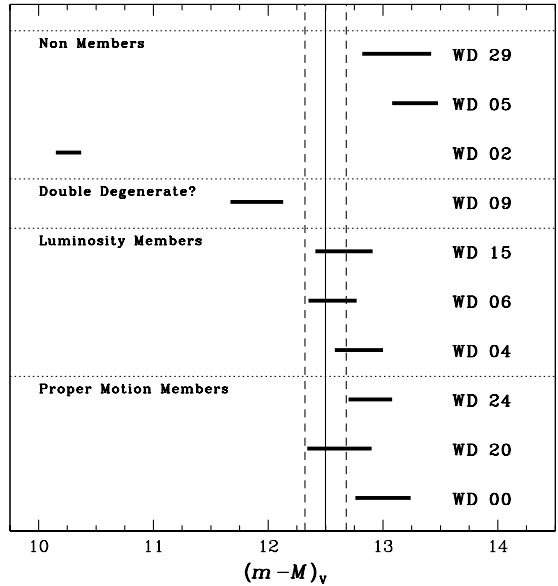} 
\figcaption{The combination of the theoretical magnitude of each white dwarf, from 
fitting the spectrum with models, and the observed magnitude is used to calculate 
individual distance moduli.  The length of the bar for each white dwarf represents 
the 1$\sigma$ error in the theoretical magnitude.  The solid line indicates the distance 
modulus of M4 \citep{richer04} of ($m-M$)$_V$ = 12.51 $\pm$ 0.09, and the dashed 
lines represent 2$\sigma$ bounds.  The three objects at the 
bottom of the diagram, WD~00, 20, and 24, are known proper motion members of M4.  A 
further proper motion member, WD~09, is overluminous and appears to be 0.7~mags closer 
than the cluster.  Likely, this star is either a field object sharing the proper motion 
of the cluster or an unresolved double degenerate.  Another three objects in the middle 
of the diagram, WD~04, 06, and 15 are all consistent with membership whereas three 
objects at the top, WD~02, 05, and 29 are excluded from our member sample.
\label{fig:mags}}
\end{figure}



We compare the calculated distance modulus of each of our ten white dwarfs 
with the known M4 distance in Figure~\ref{fig:mags}.  The 1$\sigma$ uncertainty 
in the theoretical magnitude (i.e., from spectral fitting) is plotted as a horizontal 
bar for each star, and the mean distance of M4 and $\pm$2$\sigma$ uncertainties are indicated 
with the solid and dashed lines.  We can use the proper motion members of the 
cluster as a rough guide to establish the expected scatter of known members in this 
diagnostic. The calculated distances of three of the four proper motion members of 
the cluster, WD~00, 20, and 24, are close to, but slightly larger than the distance of M4, 
whereas the fourth star (WD~09) appears to be 0.7~mags closer than the cluster.  This object could 
therefore be a non-member that shares the proper motion of the cluster.  Alternatively, 
if the object is a member, its overluminous nature may suggest an unresolved double 
degenerate in the cluster.  We note that the mass derived from 
the spectrum of this star yields the unexpectedly high value of $M$ = 0.79~$M_\odot$ 
(see Table~2).  Three additional white dwarfs are clearly members 
of the cluster, WD~04, 06, and 15, whereas three other stars WD~02, 05, and 29 are 
possibly field white dwarfs.  

To summarize, our final secure sample of singly evolved, M4 member white 
dwarfs includes three objects that are proper motion members (WD~00, 20, 
and 24) and three objects which are on the cooling sequence and have 
the correct distance from spectral fitting (WD~04, 06, and 15).

\subsection{Analysis of $M_{\rm final}$}\label{analysis}

Theory predicts that the masses of white dwarfs forming in old, population II 
systems such as globular clusters are expected to be in the range 0.51 $ < M <$ 
0.55~$M_\odot$ \citep{renzini88,renzini96}.  As Table~2 indicates, the six 
single white dwarfs that we have confirmed as members of M4 have masses that range 
from 0.50 -- 0.59~$M_\odot$.  The uncertainties in our mass measurements are 
all roughly the same, under 10\%, and therefore the un-weighted mean mass of the 
sample of six white dwarfs is measured to be $M_{\rm final}$ = 0.53 $\pm$ 
0.01~$M_\odot$.  

The present day temperature of these cooling remnants is also well measured from 
our data.  There is a group of four stars in our sample with $V$ $\sim$ 22.6 
($M_V$ = 10.1) and all have the same temperature, $T_{\rm eff} \sim$ = 25,000~K.  
The two fainter stars, WD~00 and 20, are cooler than this group by 
about 5,000~K, as expected.  The surface gravities, log~$g$, of all six stars 
are in the range log~$g$ = 7.68 -- 7.95.  As mentioned earlier, the exact 
$T_{\rm eff}$ and log~$g$ values have not been adjusted to reflect the small 
corrections required by the new Stark broadening calculations \citep{tremblay09}.

Our sampling of M4 cluster member white dwarfs in this study puts us in the fortunate 
situation of having characterized a group of four remnants at the same 
position on the CMD.  \cite{hansen07} demonstrate that the mass difference between 
white dwarfs at the tip of the cooling sequence to those five magnitudes fainter 
at the limit of the HST study is only $\sim$0.1~$M_\odot$.  These four white 
dwarfs, WD~04, 06, 15, and 24, differ in luminosity by just 0.07~mag, and therefore 
should have the same mass, and have evolved from stars with the same progenitor mass.  The 
expected temperature range in this group of four stars should only be $\sim$2\%.
As these differences are much smaller than our uncertainties, we can verify our 
mean mass measurement of M4 white dwarfs by creating a composite white dwarf spectrum, 
with very high S/N, from the coaddition of these individual spectra.  \cite{liebert91} 
have shown that for unresolved double degenerate systems, the atmospheric parameters 
derived from assuming a single star are in fact an average of the parameters of both 
components of the system.  Therefore, spectra with identical gravities should yield 
the same average gravity in a coaddition as compared to the true individual values.

Such a coaddition has the advantage that the resulting fit will be minimally influenced by 
any artifact affecting a given Balmer line in one spectrum (e.g., presence of a cosmic ray).  
This method has several disadvantages as well, for example, the spectral lines in the 
composite may be artificially blurred if the wavelength solution is not well determined 
in each of the individual spectra.  We stress that this test is only intended as a 
verification of our results and should not be attempted in samples where the mass 
variation between individual white dwarfs is non-negligible (e.g., similar studies in 
younger clusters).  

\begin{figure}
\epsscale{1.1} \plotone{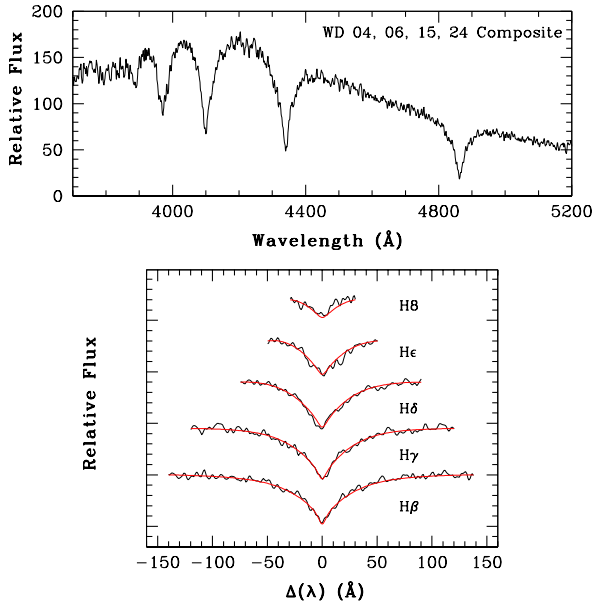} 
\figcaption{A S/N $\sim$ 100 composite spectrum of four individual M4 white dwarf 
members, all of which have the same luminosity to within 0.07~mags.  The variation in 
mass between the individual stars is expected to be negligible.  The spectroscopic 
fit to the Balmer lines of the composite spectrum yields a mass of $M$ = 0.54 
$\pm$ 0.02~$M_\odot$, in excellent agreement with our overall mean mass of the 
total confirmed sample of member white dwarfs.  As for the spectra shown in 
Figure~\ref{fig:wdspectraM4}, we have smoothed the composite spectrum by three 
pixels in this Figure.
\label{fig:composite}}
\end{figure}

The composite spectrum from this analysis is shown in Figure~\ref{fig:composite} (top), 
and exhibits a S/N $\sim$ 100 at $\lambda$ = $\sim$4030~${\rm \AA}$.  The Balmer lines 
in this spectrum are very accurately characterized; the resulting best fit model to 
the five lines is shown in the lower panel.  The spectroscopic mass from the fit is 
measured to be $M$ = 0.54 $\pm$ 0.02~$M_\odot$, in excellent agreement with our 
findings above.


Overall, from our spectroscopic study of white dwarfs in M4, we can conclude that 
the mass of population II remnants forming today is $M_{\rm final}$ = 0.53~$M_\odot$ 
and the temperature of stars presently at the tip of the white dwarf cooling sequence 
is $T_{\rm eff}$ = 25,000~K. 

\section{Stellar Evolution Theory}

\subsection{The Initial-Final Mass Relation} \label{ifmr}

The mass distribution of white dwarfs in nearby open star clusters has been 
mapped since 1977 when Volker Weidemann compared theoretical models of mass loss 
(e.g., Fusi-Pecci \& Renzini 1976) to the observed masses of a few remnants 
in the Hyades and Pleiades clusters \citep{weidemann77}.  Since this pioneering 
work, similar mass measurements have been made for over 100 stars in a 
dozen open star clusters.  A review of the earlier work is provided in 
Weidemann~(2000) and a compilation of some more recent results is presented in 
\cite{ferrario05}, excluding very recent studies by \cite{dobbie06,williams07,kalirai07,kalirai08,williams09,dobbie09}.  
The mass distribution of the remnants varies as a function of the turnoff mass of their 
parent cluster, suggesting that more massive progenitors produce more massive 
white dwarfs (see Salaris et~al.\ 2009 for a detailed analysis of the present relation).  
New measurements by our team in three older star clusters with ages of a few Gyrs have 
recently permitted, for the first time, a purely empirical fit to the data without 
the need for a theoretical anchor at low masses (see below).

\begin{figure}
\includegraphics[width=8cm]{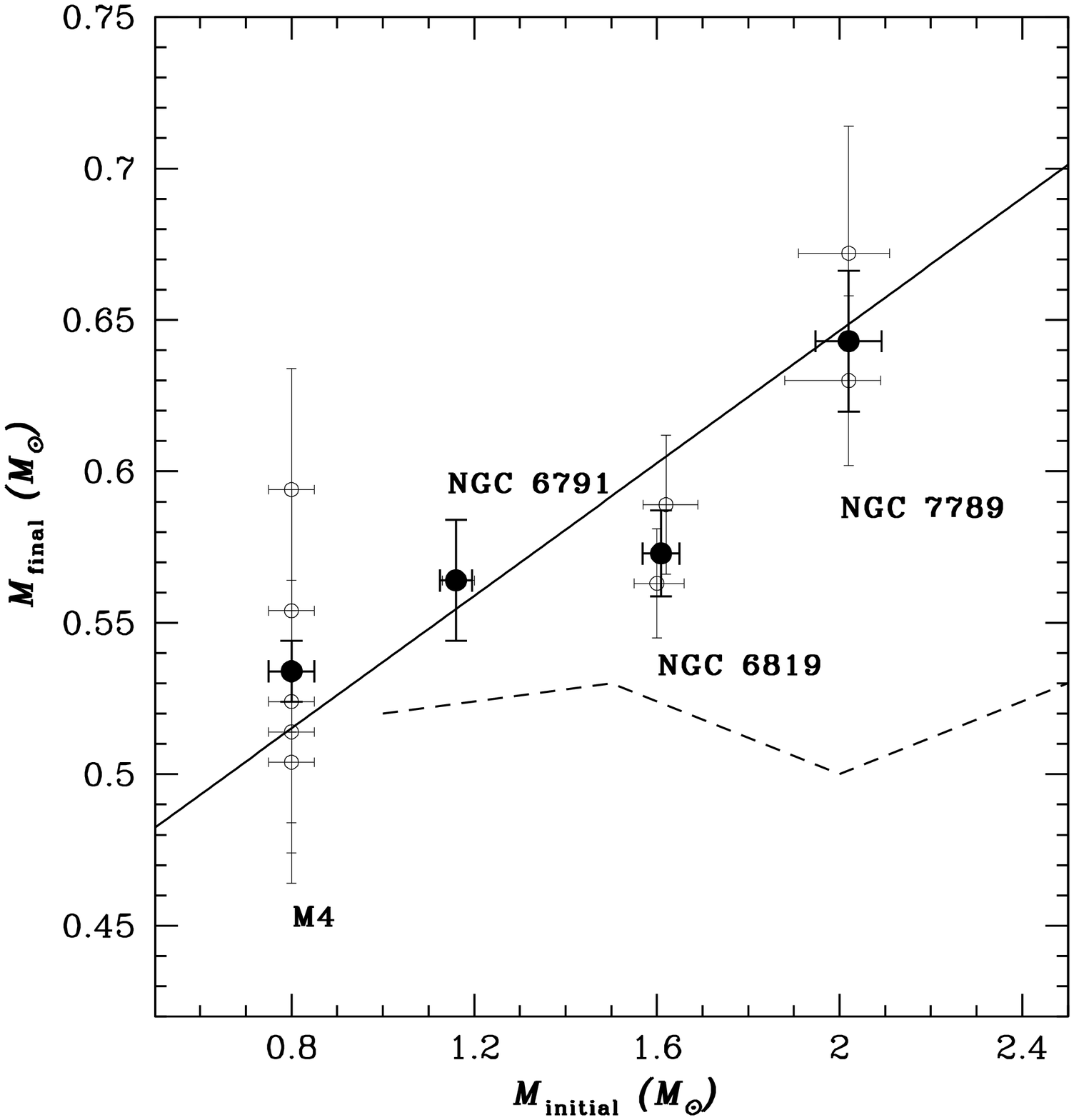}
\caption{The low mass end of the initial-final mass relation showing 
white dwarf mass measurements and calculated progenitor masses in 
NGC~7789 and NGC~6819 \citep{kalirai08}, NGC~6791 \citep{kalirai07}, 
and the present study for M4.  All white dwarf mass measurements have been 
corrected by $+$0.034~$M_\odot$ to reflect the new Stark broadening calculations 
presented in \cite{tremblay09}.  As discussed in \S\,\ref{masslossmetallicity} and 
\cite{kalirai07}, the single white dwarf in NGC~6791 is much more massive 
than the mean mass of remnants in the cluster, most of which are low mass 
helium-core white dwarfs (and not shown here).  The individual white dwarf 
-- progenitor pairs are shown as open circles and the binned averages in 
each cluster are shown as larger filled circles.  The relation shows a 
linear decline of final remnant mass with decreasing initial progenitor mass, 
extending down to the extreme low mass end, anchored by our M4 measurement at 
$M_{\rm initial}$ = 0.80 $\pm$ 0.05~$M_\odot$ and $M_{\rm final}$ = 0.53 $\pm$ 
0.01~$M_\odot$.  The solid line shows the best fit empirical relation 
to all individual data points excluding the new M4 white dwarfs, and the 
dashed curve shows the mass of the core at the first thermal pulse (see \S\,\ref{ifmr}).
\label{fig:ifmr}}
\end{figure}

With the present study, we can once again extend the initial-final mass relation to 
new bounds.  As the initial mass function in most (all?) stellar populations is bottom 
heavy, the characterization of mass loss from these low mass stars affects our 
understanding of stellar evolution for the bulk of all stars today.  In 
Figure~\ref{fig:ifmr}, we present all of the constraints to date at the low mass end 
of the relation, corrected by $+$0.034~$M_\odot$ from our previous reporting 
to reflect the new Stark broadening calculations (see \S\,\ref{modelfit}).  
The data includes the two confirmed cluster members in the 
1.4~Gyr cluster NGC~7789 \citep{kalirai08}, the two white dwarfs in the 2.5~Gyr 
cluster NGC~6819 \citep{kalirai08}, the single carbon-oxygen core white dwarf in 
the 8.5~Gyr cluster NGC~6791 \citep{kalirai07}, and the six white dwarfs in M4 
from this study.  For the latter, we have fixed the progenitor mass to the expected 
mass at the turnoff in an old, metal-poor population, $M_{\rm initial}$ = 0.80~$M_\odot$, 
with an adopted error of 0.05~$M_\odot$.\footnote{The exact turnoff mass depends on the 
metallicity and age of the cluster, and the choice of theoretical model assumed.  For 
example, for alpha-enhanced models with [Fe/H] = $-$1.1 and $t$ = 12.0~Gyr, the 
VandenBerg, Bergbusch, \& Dowler~(2006) models yield a turnoff mass of 0.82~$M_\odot$.}  
The data points for individual white dwarfs 
in these clusters are shown as open circles and binned averages of the population in 
each cluster are illustrated with larger, filled circles.  For older clusters, all of 
the white dwarfs at the top of the cooling sequence evolved from progenitors just above 
the present day turnoff, and therefore we expect a single mass at the cooling sequence 
in a given cluster.  Of course, in younger clusters, our spectroscopic measurements 
can include stars along the entire cooling sequence of the cluster and therefore 
there can be a large dispersion in white dwarf masses in a single cluster (e.g., 
NGC~2099, Kalirai et~al.\ 2005 and NGC~2168, Williams, Bolte, \& Koester 2009).  


The low mass end of the initial-final mass relation in Figure~\ref{fig:ifmr} 
now contains a total of eleven data points, and shows a roughly linear decrease in white dwarf 
mass with initial mass, anchored at the extreme left end by the new M4 measurement, 
$M_{\rm initial}$ = 0.80 $\pm$ 0.05~$M_\odot$ and $M_{\rm final}$ = 0.53 $\pm$ 
0.01~$M_\odot$.  Given that the oldest observable stellar populations in the Universe 
are the same age as M4, this relation is now complete at the lower limit 
and can not be observationally extended further to lower masses.  As discussed earlier, 
several predictions of stellar evolution theory have long suggested that the masses of 
population II white dwarfs forming today should be 0.51 $ < M_{\rm final} <$ 0.55~$M_\odot$ 
\citep{renzini88,renzini96}, and therefore our first direct measurement is in 
perfect agreement with these predictions.  Based on our measurements, the total 
integrated post main-sequence mass loss of low mass stars in old population II systems 
will be $\sim$34\%.  

Theoretically, there is an expected lower limit to the masses of carbon-oxygen core white 
dwarfs.  This is given by the mass of the stellar core at the first thermal pulse on 
the asymptotic giant branch, an evolutionary point before the core has had a chance to 
grow on the branch.  This core mass has a very small dependency on the stellar metallicity.  
In Figure~\ref{fig:ifmr}, we illustrate this expected core mass as a dashed curve 
\citep{girardi00}.  Over the range of initial masses on this plot, the core mass is 
expected to remain roughly the same at $M_{\rm final}$ = 0.52~$M_\odot$, showing a flat relation 
for $M_{\rm initial}$ $\sim$ 1 -- 2~$M_\odot$.  At 0.8~$M_\odot$, our first observational 
constraint from a half dozen white dwarfs in this single star cluster 
suggests the actual remnant masses are just slightly larger than the minimum core mass 
(with a slight extrapolation).  This implies that the progenitors of such low mass stars had 
just enough mass to experience thermal pulses on the asymptotic giant branch and that the cores 
of these stars will not grow appreciably during this evolution.  As M4 is a typical 
population~II system in terms of its old age and low metallicity, it is reasonable to 
assume that these trends may extend to stars that are presently in similar phases of 
stellar evolution both in the general field halo population and in other globular 
clusters with similar properties to M4.

\subsection{The Dependence of Mass Loss on Metallicity} \label{masslossmetallicity}

In addition to extending the initial-final mass mapping to lower masses, M4 
represents the first star cluster with a significantly sub-solar metallicity in 
which white dwarf masses have been measured.  The mean metallicity of the cluster 
is [Fe/H] = $-$1.10 \citep{marino08}, 13 times more metal-poor than the 
bulk of the open clusters used to define the relation and 40 times more metal-poor 
than the metal-rich system NGC~6791 ([Fe/H] = $+$0.40, Peterson \& Green 1998).  For 
this cluster, after 
accounting for the new Stark broadening calculations discussed earlier, \cite{kalirai07} 
actually measured a mean mass for nine white dwarfs of $M_{\rm final}$ = 
0.46 $\pm$ 0.06 $M_\odot$, much lower than expected.  As they show, the bulk of these 
white dwarfs contain helium-cores, after evolving through a channel involving intense 
mass loss on the red giant branch, likely driven by the high metallicity of the cluster.  
The single data point shown in Figure~\ref{fig:ifmr} is based on the only clearly non 
He-core white dwarf in that sample, and does not affect the discussion above.  However, 
the results from NGC~6791 suggest strong dependencies of mass loss on metallicity, 
especially at super-solar metallicities.  

To date, we lack a good theoretical understanding of the dependence of post main-sequence 
mass loss on metallicity.  The general expectation in evolutionary models is for more 
efficient transport of material outward from the star if the metallicity is higher, 
and therefore higher rates of mass loss.  However, the exact treatment depends on the 
modeling of winds on the red giant branch and thermal pulses on the asymptotic giant branch 
(e.g., see Weidemann~2000 and Habing 1996), phases of stellar evolution that are poorly 
understood.  As \cite{catelan08} summarizes, mass loss prescriptions along the red 
giant branch alone show very different behaviors with metallicity.  For example, 
over a 1~dex metallicity change from [Fe/H] = $-$1.0 to 0.0, the \cite{mullan78} and 
\cite{schroder05} laws predict that $\Delta$$M_{\rm RGB}$ should increase by $\lesssim$0.05~$M_\odot$ 
whereas the \cite{goldberg79} and \cite{judge91} laws suggest $\Delta$$M$ increases by 
0.15 -- 0.25~$M_\odot$.

To qualitatively compare the dependence of mass loss in post main-sequence evolution with 
metallicity to these estimates, we can first parameterize the initial-final mass relation.  
The solid line in Figure~\ref{fig:ifmr} is a linear fit to the entire sample of {\it open cluster} 
white dwarf -- main-sequence progenitor mass pairs from \cite{kalirai08},

\begin{eqnarray*}
M_{\rm final} = (0.109 \pm 0.007)~M_{\rm initial} + 0.428 \pm 0.025~M_\odot. \\
\end{eqnarray*}

The zero point of this relation has been adjusted to reflect a global offset of 
$+$0.034~$M_\odot$ in the white dwarf mass measurements, although the slope may in fact 
also be affected if the offset is found to vary with mass (see Figure~11 and 12 in 
Tremblay \& Bergeron 2009).  The bounds on this relation 
range from the oldest open cluster white dwarf in NGC~6791 to the 
youngest white dwarfs in the Pleiades, and therefore are $M_{\rm initial}$ = 1.1 -- 6.5~$M_\odot$.  
A small extrapolation of the empirical relation to the mass of the present day turnoff in M4 
(e.g., $\sim$0.8~$M_\odot$) yields $M_{\rm final}$ = 0.52 $\pm$ 0.02~$M_\odot$, which is 
therefore consistent (at 1$\sigma$) with our measured value of the masses of M4's white dwarf 
population.  These data therefore {\it hint} that mass loss rates have a weak or no dependence on 
metallicity, at least over the range extending from metal-poor populations such as M4 to 
roughly solar metallicity.  Interestingly, we know from our study of 
NGC~6791 that mass loss rates are strongly correlated with metallicity at extremely high [Fe/H], 
and therefore this may lead to an interesting shape for the empirical mass loss -- metallicity 
relation from white dwarf spectroscopy.  Of course, stronger conclusions will benefit greatly from 
additional observations of clusters spanning a wide range of metallicities, especially at the 
metal-poor end.  For completion, if we add our new data point for M4 to the initial-final mass 
relation and recalculate the best fit ignoring any possible metallicity-related biasses, 
we find a slightly shallower relation with $M_{\rm final}$ = 
(0.101 $\pm$ 0.006)~$M_{\rm initial}$ + 0.463 $\pm$ 0.018~$M_\odot$.

Finally, we note that the metallicity of M4, [Fe/H] = $-$1.10, is only slightly more metal-rich
than the global metallicity of both the stellar halo of the Milky Way ([Fe/H] = $-$1.6, 
$\sigma$ = 0.6; Morrison et~al.\ 2003) and M31 ([Fe/H] = $-$1.5, $\sigma$ = 0.7; Kalirai et~al.\ 
2006).  Assuming the dominant population in these galactic components is old (e.g., 
$\gtrsim$10~Gyr), the same 34\% stellar mass loss fraction should be representative of stars 
evolving in these populations.  By directly inputing this mass loss into general stellar 
evolution models of similar age and metallicity (e.g., Girardi et~al. 2000), we can add an 
important observational constraint to the modeling of light from the stellar halos of spiral 
galaxies using population synthesis techniques (e.g., Bruzual \& Charlot 2003).  

\section{The Ages and Distances of Globular Clusters}\label{WDCoolingAges}

In Section~\ref{introduction} we introduced the extensive HST observations that 
have defined some of the deepest CMDs ever measured, with any instrument 
(e.g., the HST Cycle 13 observations of NGC~6397 in Richer et~al.\ 2008).  
Although these data sets have led to a wide range of astrophysical studies, the 
primary goal of the observations has aimed to accurately establish turnoff-independent 
white dwarf cooling ages for population II star clusters.  In this respect, the 
present study serves to establish several of the input ingredients that are needed 
to make this measurement.  For example, knowledge of the masses and spectral types 
of white dwarfs, and the initial-final mass relation, are all required to fully 
model an observed white dwarf cooling sequence (e.g., Hansen et~al.\ 2004; 2007).

\subsection{Specific Ingredients for Accurate Age Measurements}

For M4, two recent studies have modeled the white dwarf cooling sequence 
using independent data sets, and both have concluded with similar ages (however, see discussion 
below).  First, \cite{hansen04} used a full 2-D treatment of the white dwarf cooling sequence 
in the observed HST WFPC2 CMD (i.e., luminosity and color information are used), and just 
recently, \cite{bedin09} have used a more simplified technique of age dating based solely on 
the luminosity function in newer and deeper ACS data.  An analysis of these studies 
illuminates several interesting choices relative to the findings in this paper:

\begin{itemize}

\item The present day mass of cluster white dwarfs forming today sets the 
normalization for the masses along the entire cooling sequence.  In their modeling, 
\cite{hansen04} assume the mass at the tip to be 0.55~$M_\odot$ and \cite{bedin09} 
assume 0.54~$M_\odot$.  As we have shown, the actual mass of white dwarfs at the 
bright end of the cooling sequence is only slightly lower at 0.53~$M_\odot$, and 
therefore the mass distribution (and hence cooling rates) along the entire cooling 
sequence are not very different from the assumptions in these studies.  

\item Hydrogen (DA) and helium (DB) atmosphere white dwarfs have different cooling rates, 
and so the fraction of stars cooling with these spectral types leads to structure in the 
luminosity and color function that must be modeled correctly to infer the age of the 
population from white dwarf cooling theory.  For M4, NGC~6397, and NGC~6752, spectroscopy 
of the white dwarf cooling sequence indicates that 100\% of the stars at the bright end have 
hydrogen atmospheres (see Moehler et~al.\ 2004 and Davis et~al.\ 2009).  \cite{hansen04} 
treated the fraction of DB white dwarfs as a free parameter and found the best fit to 
yield an upper limit of 40\% (good fits were obtained as long as more than 60\% of the 
white dwarfs are DA), whereas \cite{bedin09} choose an arbitrary value of 30\% (e.g., 
the field disk ratio).

\item The initial-final mass relation is required to properly evolve the progenitor 
stars on to the white dwarf cooling sequence.  The normalization and form of the 
initial-final mass relation adopted in both of these studies (e.g., Wood 1992), is 
based on empirical constraints from solar metallicity data points at higher masses, 
and a theoretical anchor at low masses.  This can now be updated to our new result 
given earlier.

\end{itemize}

We encourage future theoretical modeling efforts to make full use of this new 
information in their analysis, preferably on the newer ACS imaging data of M4.
Of course, the white dwarf models themselves should also be updated with new physical 
insights.  For example, the contribution to the opacity of cool white dwarfs from 
the far red wing of the Ly$\alpha$ line (Kowalski 2007) alone can yield ages of 
globular clusters (e.g., NGC~6397) that are higher by $\sim$0.5~Gyr (Kowalski 2007; 
B.\ Hansen 2009, private communication).

\subsection{A Closer Look at M4's White Dwarf Cooling Age: \\ Model and Distance Uncertainties}

We noted above that the \cite{hansen04} and \cite{bedin09} studies concluded 
with similar white dwarf cooling ages for M4.  Formally, \cite{hansen04} reported 
an age of 12.1~Gyr (2$\sigma$ lower limit of 10.3~Gyr) and \cite{bedin09} measured 
an age of 11.6 $\pm$ 0.6~Gyr (internal errors).  A deeper look at these two studies 
suggests they are, in fact, grossly {\it inconsistent} with one another given 
differences in white dwarf cooling models and uncertainties in the cluster distance 
modulus. As our present work is closely linked to obtaining very accurate absolute 
age measurements of globular clusters (through the study of white dwarfs), we briefly 
summarize these two studies below and highlight the respective assumptions and 
differences that are being made. 

In the \cite{hansen04} study, the authors compare their data to three sets of 
models and adopt a distance to M4 based on the well established technique of 
subdwarf distance fitting.  They find $d$ = 1.73 $\pm$ 0.10~kpc for M4 (e.g., 
Richer et~al.\ 1997), which is identical to independent distance measurements for 
the cluster from the Baade-Wesselink distance using the M4 RR Lyrae stars 
($d$ = 1.73 $\pm$ 0.01~kpc, Liu \& Janes 1990) and from astrometry ($d$ = 1.72 
$\pm$ 0.14~kpc, Peterson et~al.\ 1995).  At present, it appears the latter estimate 
is the most secure.  For the subdwarf measurement, the comparison was based on 
pre-Hipparcos parallaxes, and for some other clusters, a calculation with 
post-Hipparcos parallaxes indicates a longer distance by $\sim$0.2~mags 
than initially estimated (Gratton et~al.\ 1997).  Similarly, the 
Baade-Wesselink method yields luminosities for M4's RR Lyrae stars that are 
fainter by $\sim$0.2~mags as compared to the $M_V$--[Fe/H] relation derived by 
Gratton et~al.\ (1997) for clusters with low extinction (using post-Hipparcos 
main-sequence fitting techniques).  Unfortunately, re-deriving the distance to 
M4 using any photometric technique is problematic given the very high extinction 
along the line of sight ($A_V$ $>$ 1), the variation in $A_V$ across the face of 
the cluster, and the non-standard shape of the extinction curve along this line 
of sight (which passes through the Scorpius-Ophiucus dark cloud complex -- Vrba, 
Coyne, \& Tapia 1993).  Irrespective of the adopted distance (although 
see below), \cite{hansen04} find that the white dwarf cooling age they derive for 
M4 from their own models (e.g., Hansen 1999) is consistent with that measured using 
the \cite{chabrier00} models, however they rule out the \cite{salaris00} models 
which yield ages that are older than the Universe.  Therefore, this suggests 
that, for the same distance modulus and data set, simply changing the white dwarf 
model yields an absolute age difference of 2~Gyr.

In the recent \cite{bedin09} study, the authors use only the \cite{salaris00} white dwarf 
models, and still conclude with an age similar to that derived by \cite{hansen04}.  The 
difference is actually reflected in the distance to M4; \cite{bedin09} use a much 
larger (i.e., 15\%) value of $d$ = 1.98~kpc.  This value is formally inconsistent with 
the three independent studies above.  \cite{bedin09} measure the distance to M4 by 
simply fitting a single isochrone of approximately the correct metallicity to the turnoff and 
horizontal branch.  In principle, this method can of course yield the distance to any 
stellar population with a measured turnoff, however it has almost always been avoided 
for a large number of reasons.  First, and most importantly, there are uncertainties in 
both the microphysics built into the stellar evolution models (e.g., opacities, equation of 
state effects, and nuclear reaction rates) and in the treatment of processes that do 
not come from first principles (e.g., convection, core-rotation, diffusion, gravitational 
settling, etc.).\footnote{\cite{vandenberg96} present a detailed discussion 
of these effects on stellar evolution models.}  Because of these unknowns, simply 
adopting an isochrone from another groups' models would lead to a different measured 
distance modulus for the cluster (e.g., different models predict different horizontal branch 
luminosities).  Second, such comparisons of data and models to yield distances require 
transformation of the model from the theoretical to the observational plane, resulting 
in another source of error.  Finally, in this specific case, this comparison does not 
alleviate any of the aforementioned uncertainties related to our lack of knowledge of 
the reddening and extinction along this line of sight.  In the end, the application of 
the \cite{salaris00} models and the longer distance modulus derived by \cite{bedin09} 
effectively offset each other, yielding a derived age similar to that of \cite{hansen04}. 

Summarizing, there are clearly both intrinsic differences in white dwarf models calculated 
by different groups and there is a lack of consensus on the correct distance to M4.  
For the former, increased testing of the white dwarfs models on star clusters with well 
known fundamental parameters (e.g., distance, metallicity, and foreground reddening) are 
sure to resolve the correct prescriptions.  For the latter, it appears that only the 
\cite{peterson95} astrometric measurement is free of the unknown reddening and extinction 
laws along this line of sight, and therefore should be preferred.  Until these differences 
are unambiguously resolved (e.g., the Space Interferometry Mission could directly measure the 
distance to M4 to within a few percent), the error budget from these uncertainties will 
remain large and the new physical constraints on masses, spectral types, and the initial-final 
mass relation are secondary to the analysis.  Of course, if the properties of bright white 
dwarfs in other population II systems are similar to M4, the new results can have immediate 
significance on those studies.  This is especially exciting given the remarkable data set 
that has been presented for NGC~6397 by \cite{richer08} and will soon be available for 
47~Tuc (H.\ Richer, Cycle 17 ACS/WFC3 Proposal GO-11677). 



\subsection{White Dwarfs as Standard Candles} \label{candles}

The mass measurement of white dwarfs in a globular cluster provides important leverage to the 
use of these stars as distance calibrators.  Although the white dwarf cooling sequence is 
fainter than the bright part of the main-sequence in a star cluster, uncovering the sequence 
can offer several advantages over canonical techniques involving main-sequence distance 
fitting.  For example, the location of the bulk of the white dwarf cooling sequence in a 
CMD does not depend on the metallicity or age of the population, or on any of the other 
systematics that plague main-sequence evolution (e.g., convection theory).  However, the 
location does depend on mass (e.g., Wood~1995), with a 0.1~$M_\odot$ difference translating 
to a $\Delta$(mag) = 0.25 shift (i.e., a 12\% distance offset).  So, first, a simple offset 
can now be calculated to correct the location of any local calibrating sample of white dwarfs 
with parallax measurements (usually 0.60~$M_\odot$ for field white dwarfs) to the expected 
location of 0.53~$M_\odot$ stars.  Following this, a straightforward comparison with the 
apparent luminosity of the globular cluster white dwarf cooling sequence can yield the 
distance to the population (e.g., see Renzini et~al.\ 1996).  Unfortunately, it is 
difficult to actually make this measurement and independently derive the distance of M4 
given the aforementioned uncertainties in the slope of the extinction curve along this 
line of sight.  However, the true distances of other clusters with well constrained 
reddenings and extinctions can be easily calculated from their cooling sequences and 
the new mass measurements.

\section{Summary}

We have presented the first directly determined spectroscopic mass measurements of 
white dwarfs that are known to be population II members.  Our measurements 
suggest that the final remnant mass at the tip of the white dwarf 
cooling sequence in the globular cluster M4 is $M_{\rm final}$ = 0.53 $\pm$ 
0.01~$M_\odot$, in nice agreement with expectations based on stellar evolution 
theory.  These results suggest that low mass population II stars will lose 
$\sim$34\% of their mass through post main-sequence stellar evolution.  If, 
in fact, all Milky Way globular clusters have the same age, then this mass loss 
rate should be applicable to the evolution of stars in all but the most metal-rich 
Galactic globular clusters.  We use this measurement to first extend 
the initial-final mass relation of stars to the lowest possible limit, given the age of 
the Universe.  This single observation also hints that mass loss rates are not 
significantly different for metal poor stars when compared to solar metallicity stars 
of the same mass.  Finally, we offer several pieces of new input that are required 
to measure more accurate white dwarf cooling ages of globular clusters in the 
Galactic halo.

Imaging observations of globular clusters with ground and space based telescopes have 
now uncovered thousands of white dwarfs in population II clusters (e.g., see 
Figure~\ref{fig:globsWDs}).  The science that results from these purely photometric 
surveys is greatly enhanced with spectroscopic follow up, which we have shown can yield 
the spectral types of the remnants \citep{davis09}, the mapping of initial to final mass, 
and the mass at the tip of the observed cooling sequence (and therefore a calibration for 
the masses along the entire cooling sequence).  All of the nearest globular clusters are 
in the southern hemisphere, and given their larger distances compared to M4, an extension 
of this work to these systems will require superior blue sensitivity and throughput, and 
much larger allocations of spectroscopic time on 8-meter class telescopes (e.g., see Moehler 
et~al.\ 2004).  Future white dwarf spectroscopy in these clusters will benefit greatly from 
a 30-m telescope with a multiobject spectrograph with the bluest sensitivity.  Such 
a telescope would not only allow a direct measurement of the mass loss rates in an expanded 
set of open and globular clusters, but would permit the first initial-final mass relations 
to be built from observations of a set of white dwarfs in a {\it single} cluster.  
Deep spectroscopy of white dwarfs over a range in luminosity (and therefore mass) can be 
engaged in a given cluster, minimizing the systematics introduced by metallicity variations 
in samples of clusters.  Fortunately, the nearest three globular clusters also sample a 
wide range in metallicity (e.g., a factor of 20 from NGC~6397 at [Fe/H] = $-$2.0 to 47~Tuc at 
[Fe/H] = $-$0.7), and therefore the dependence of mass loss on metallicity can be probed 
in detail with such observations.


\acknowledgements
We wish to thank D.\ Reitzel for help with designing the spectroscopic masks in this program.  
We also wish to thank J.\ Hurley for useful discussions related to stellar evolution, 
white dwarfs, and binarity.  

JSK's research is supported in part by a grant from the STScI Director's Discretionary Research 
Fund.  The research of HBR is supported by grants from the Natural Sciences and Engineering 
Research Council of Canada. He also thanks the Canada-US Fulbright Program for the award 
of a Fulbright Fellowship.  PB is supported in part by the NSERC Canada, by the Fund FCAR 
(Qu\'ebec), and is a Cottrell Scholar of Research Corporation for Science Advancement.  
Support for MC is provided by Proyecto Basal PFB-06/2007, by FONDAP Centro de Astrof\'{i}sica 
15010003, by Proyecto FONDECYT Regular \#1071002, and by a John Simon Guggenheim Memorial 
Foundation Fellowship.  RMR acknowledges support from grant AST-0709479 from the National 
Science Foundation.


\clearpage

\appendix
\section{Snapshot Figures of All White Dwarf Candidates}



\begin{figure*}[ht]
\begin{center}
\leavevmode 
\includegraphics[width=17.5cm, bb = 18 84 592 690]{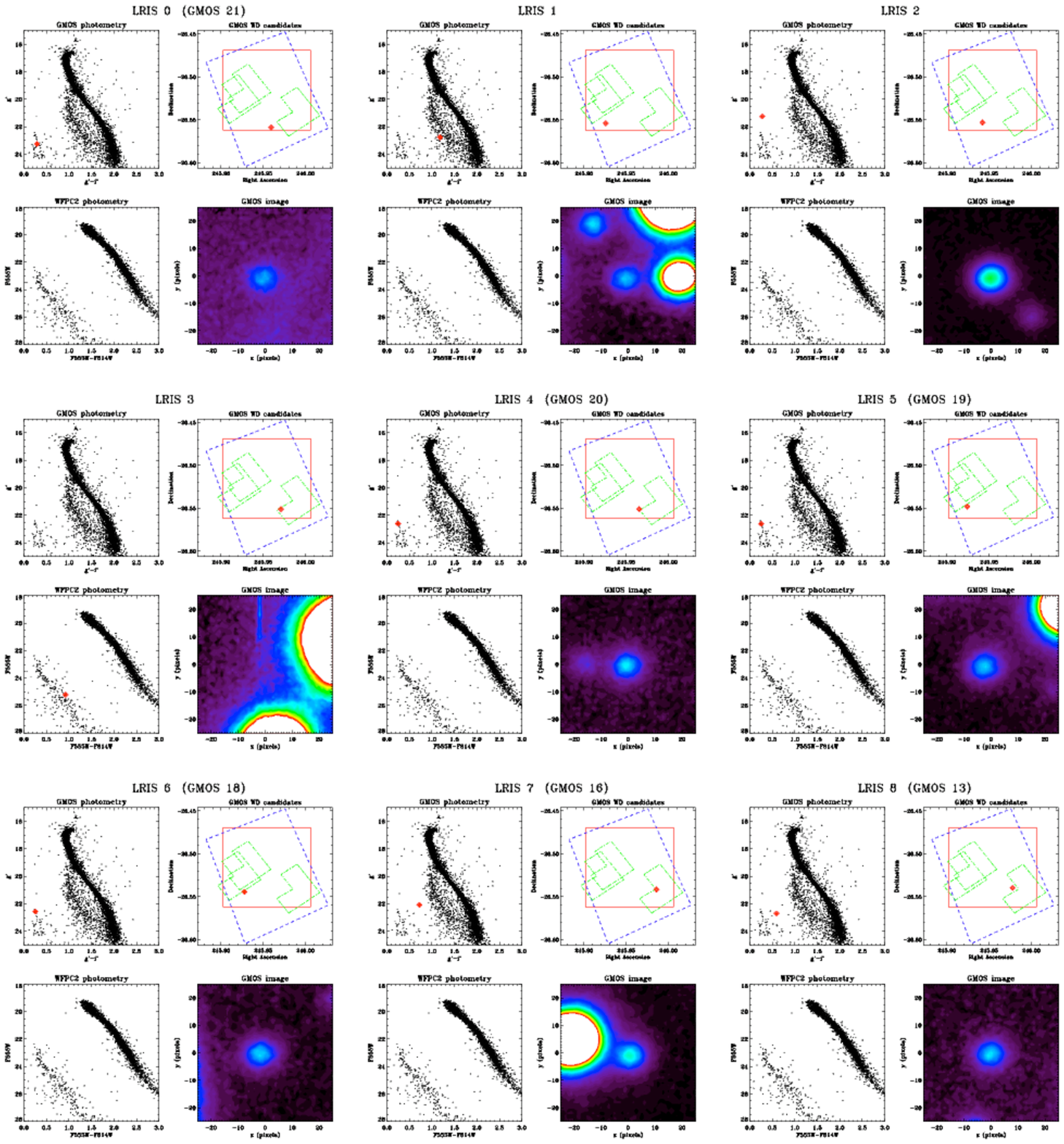}
\end{center}
\caption{These snapshot figures show several pieces of information for each of the 
white dwarf candidates WD~00 -- WD~08.  The two panels on the left illustrate the 
location of these stars on the ground based CMD (top) and the HST CMD (bottom), 
where available.  The individual candidate star is marked in each panel with a 
larger red diamond.  The location of the star relative to our imaging field of view 
is also shown in the top-right panel (see Figure~\ref{fig:foot} for details on this 
plot).  In the bottom-right panel, we present a small snapshot image of the candidate.  
Several objects that are targeted clearly have very little flux in the ground based image 
and/or are located close to bright stars.  These stars are HST detections and are discussed, 
and removed from subsequent analysis, in \S\,\ref{culling}.  Note, the identifications of 
``LRIS'' in this figure correspond to ``WD'' in Table~1.  {\it A higher resolution 
version of this figure is available in ApJ.} \label{fig:wd00_08}}
\end{figure*}



\clearpage

\begin{figure*}[ht]
\begin{center}
\leavevmode 
\includegraphics[width=17.5cm, bb = 18 84 592 710]{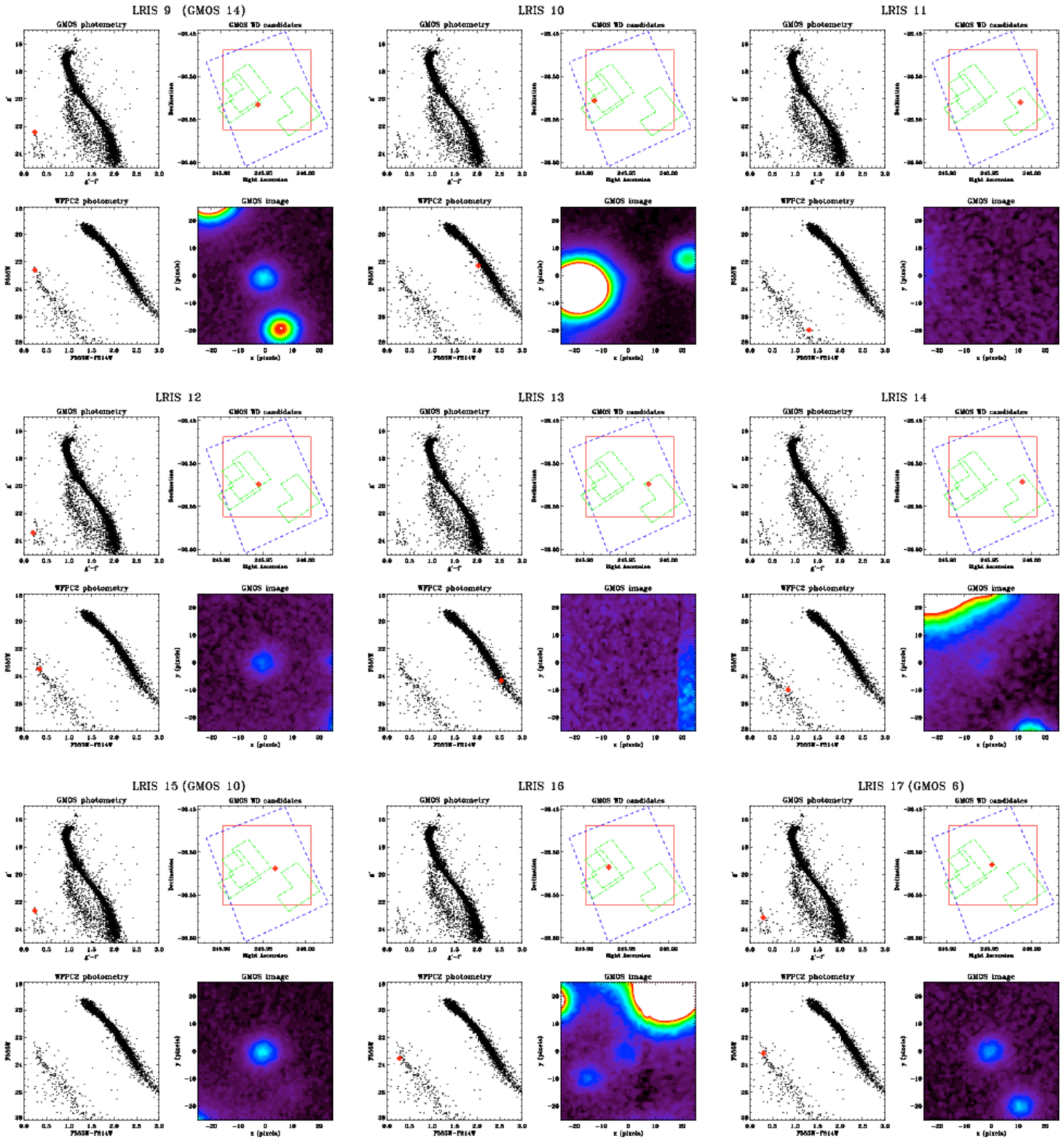}
\end{center}
\caption{Same as Figure~A1 for WD~09 -- WD~17.  {\it A higher 
resolution version of this figure is available in ApJ.}
\label{fig:wd09_17}}
\end{figure*}


\clearpage

\begin{figure*}[ht]
\begin{center}
\leavevmode 
\includegraphics[width=17.5cm, bb = 18 84 592 710]{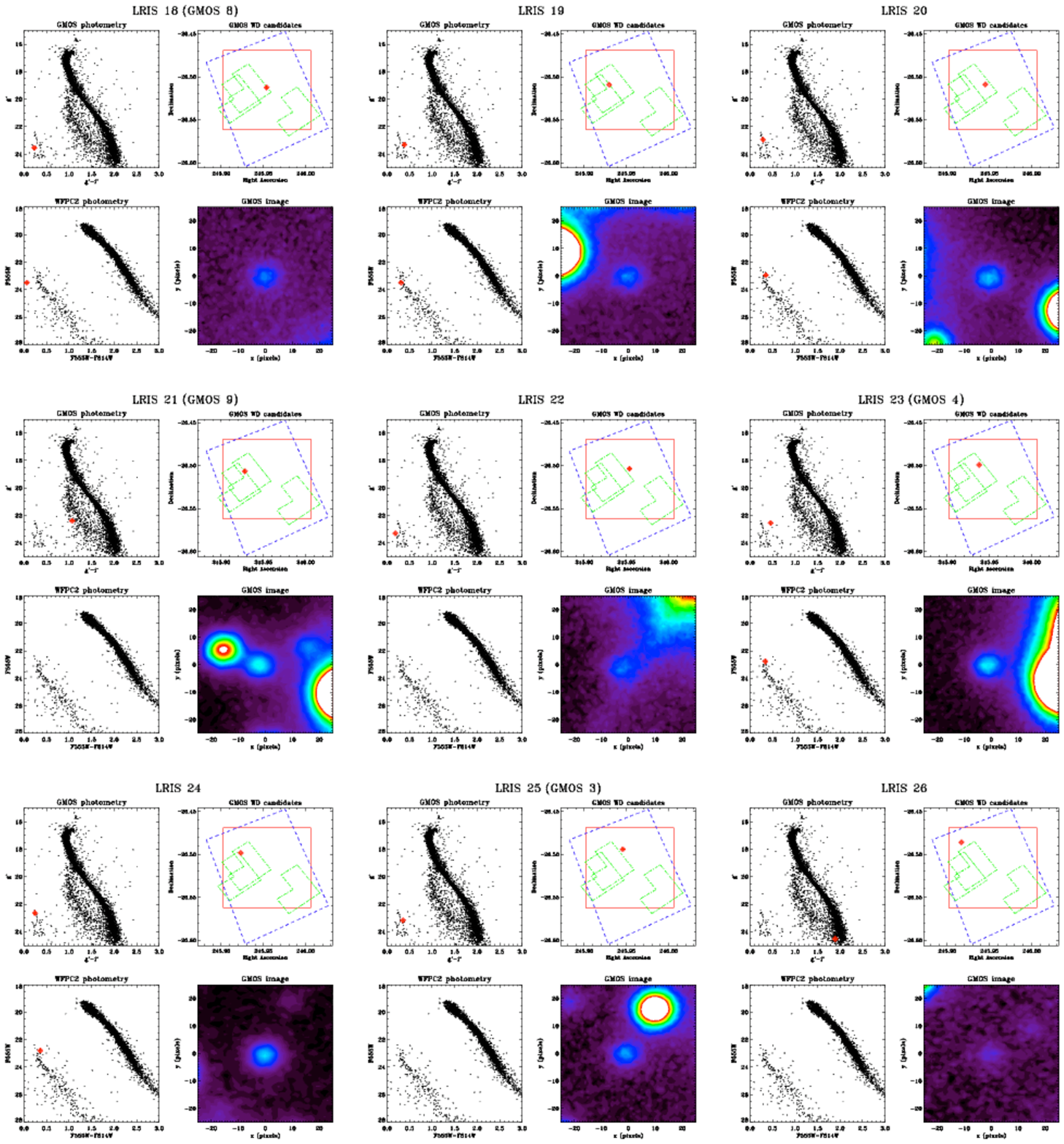}
\end{center}
\caption{Same as Figure~A1 for WD~18 -- WD~26.  {\it A higher 
resolution version of this figure is available in ApJ.}
\label{fig:wd18_26}}
\end{figure*}


\clearpage

\begin{figure*}[ht]
\begin{center}
\leavevmode 
\includegraphics[width=17.5cm, bb = 18 84 592 710]{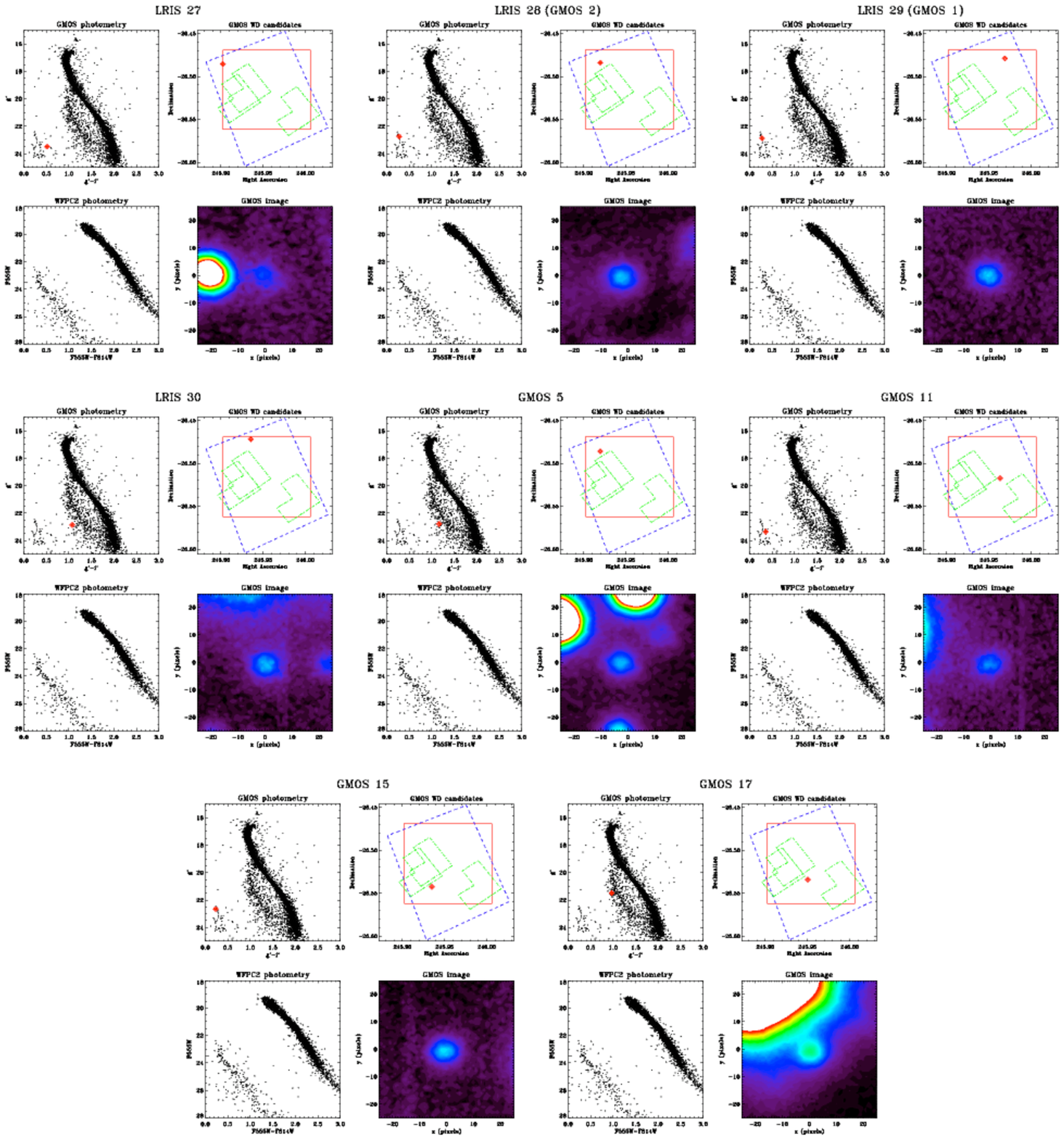}
\end{center}
\caption{Same as Figure~A1 for WD~27 -- WD~30, and GemWD~05, 
11, 15, and 17.  {\it A higher resolution version of this figure 
is available in ApJ.}
\label{fig:wd27_g17}}
\end{figure*}


\end{document}